\newcommand{\beq}{\begin{equation}}
\newcommand{\eeq}{\end{equation}}
\newcommand{\Xmax}{X^{\mu}_{\rm max}}
\newcommand{\gcm}{{\rm g\,cm^{-2}}}

\newcommand{\QII}{QGSJET-II.03}
\newcommand{\SYB}{SIBYLL2.1}
\newcommand{\EPOS}{EPOS1.99}

\documentclass[final,3p,times,12pt]{elsarticle}

\usepackage{amssymb}

 \usepackage{lineno}

\makeatletter
\def\ps@pprintTitle{%
 \let\@oddhead\@empty
 \let\@evenhead\@empty
 \def\@oddfoot{\reset@font\hfil\thepage\hfil}
 \let\@evenfoot\@oddfoot
}

\journal{ }

\begin{document}

\begin{frontmatter}

\title{A model for the transport of muons in extensive air showers}

\author[LIP]{L. Cazon\corref{cor1}}
\ead{cazon@lip.pt}
\cortext[cor1]{Corresponding author  {\it Tel} +351 217973880 {\it Fax} +351 217934631}
\author[LIP]{R. Concei\c{c}\~{a}o}
\author[LIP,IST]{M. Pimenta}
\author[LIP]{E. Santos}
\address[LIP]{LIP, Av. Elias Garcia, 14-1, 1000-149 Lisboa, Portugal}
\address[IST]{Departamento de F\'{i}sica, IST, Av. Rovisco Pais, 1049-001 Lisboa, Portugal}

\begin{abstract}
In this article we identify the key elements that govern the propagation of muons from the production in extensive air showers to ground. We describe a model based on simple assumptions that propagates the muons starting from the few relevant distributions at production. We compare the results to the ground distributions given by a full air shower Monte Carlo. This study is motivated by the need of modeling the muon component in extensive air showers with the goal of experimentally reconstructing their distributions at production, which act as a footprint of the hadronic cascade.
\end{abstract}


\end{frontmatter}


\section{Introduction}
\label{s:int}

The nature of the Ultra High Energy Cosmic Rays remains unknown. The state of the art experiments have not yet understood key aspects necessary to answer this question.  While the energy and arrival direction of the cosmic rays to Earth can be fairly well reconstructed, the primary mass is difficult to determine. The high energy spectrum may be fitted by a number of combinations of light or heavy nuclei, since the density evolution and maximum energy achievable at the sources are yet unkown. The observation of anisotropies on the  cosmic rays sky does not necessarily favor either light or heavy nuclei, because the galactic and extragalactic magnetic fields and the location of the sources are still unknown.
 The direct mass determination from air shower observables is not conclusive either. Our understanding of the hadronic particle physics is only supported up to the LHC energies. Beyond those energies  we must rely on the extrapolations of the hadronic interactions models which diverge from one to another. Besides, these new kinematic regions might uncover new phenomena not yet accommodated in models. Changes in the hadronic physics and in the composition of the primary share a region of the phase space, being difficult to break the degeneracy and answer both questions.  

Recent results from the Pierre Auger Observatory \cite{Abraham:2010yv} on the evolution of the depth of the electromagnetic shower maximum have been usually interpreted as a change towards heavier composition at the highest energies, provided that the extrapolations of the hadronic interaction models are correct. 
Nevertheless, an abrupt change on the hadronic interactions at the highest energies could be possible, leading to a rapid increase of the cross section \cite{Conceicao:2011vn} and changes on other aspects of the multiparticle hadronic production. Moreover, recent results from the LHC indicate that the current understanding of the forward direction embedded on the hadronic interactions models might be insufficient \cite{d'Enterria:2011kw,Tricomi:2010zz}.

Auger has also shown \cite{Allen:2011pe} that the number of muons in Extensive Air Showers (EAS) is underestimated by the current hadronic interactions models, even for the case of iron primaries by 40\% with respect to QGSJET-II \cite{Ostapchenko:2005nj,Ostapchenko:2006vr}. 
Muons in extensive air showers have been already the subject of many experimental studies, from their absolute number at ground to the longitudinal profile at different energies.  A book by Grieder \cite{Grieder} contains an excellent compilation of most available results. 
Very recently, the KASCADE-Grande collaboration has reported the measurement of the longitudinal profile of the production points of muons by tracking the trajectories of the detected muons at ground back to the shower axis \cite{Apel:2011zz}. They have also published the number of muons at ground for showers with different electron richness \cite{Apel:2011mi}. KASCADE-Grande reaches up to $10^{18}$ eV.

Despite of being a detector that was not originally optimized for muon reconstruction, the Auger Collaboration has recently measured \cite{GarciaGamez:2011pe} the maximum of the muon production depth profiles at energies above $10^{19.2}$ eV by mapping the arrival time of muons far from the core onto muon production distances \cite{Cazon:2004zx}. 

Recent plans from Auger include the deployment of a set of buried detectors aimed to the muonic component \cite{Alfaro:2010zz}, allowing us to explore with low systematics the region below $10^{18}$ eV, which overlaps with KASCADE-Grande and also with the LHC results.

This work is motivated by the need to extract complementary information carried by the muons, truly hadronic messengers in EAS. We aim to properly model the mechanisms that govern the muon distributions in air showers in order to peer into the details of the hadronic shower.

EAS develop in a complex way as a hadronic multiparticle production that generates a hadronic and an electromagnetic cascade which travel down the atmosphere. The electromagnetic (EM) component spreads out in time and space, reaching more than 1 km away from the shower core in large numbers, enough for detectors placed at ground level (several square meters of surface) to record their signal.  On the other hand, given their large density near the core, they produce sizable amounts of fluorescence light that can be detected with far away ultraviolet telescopes, recording their longitudinal development in moonless nights.

 The hadronic cascade is much less numerous and thus hard to detect directly. It consists mainly of low energy pions, and fewer high energetic particles, such as leading baryons and mesons that carry a large fraction of the primary energy deep into the atmosphere. The energy and momentum of these leading particles depend on the details of the high energy hadronic interaction models, which determine the production of the lower energy bulk of mesons through the inelasticity and multiplicity of the interactions.
 The hadronic cascade is the main engine of the air shower: it feeds the EM cascade mainly by the decay of neutral pions and also feeds back through the interaction of charged pions, or by means of the less numerous kaons. When a pion or kaon decays into a muon, the muon might leave the hadronic core and transports information far away from the central region.

 In \cite{a:TimeModel,a:MPD}, and later updated in \cite{a:PhDCazon}, it was shown that
the arrival time distributions of muons at ground emerge as a direct transformation of 
the muon production depth distribution and the energy spectrum at ground.
In the present paper we develop the model that links all the relevant distributions at the moment of production, namely, the muon energy, transverse momentum and production depth to the observed muon distributions at ground, namely the energy spectrum, the arrival time delay distribution, the apparent production depth distribution and the lateral distribution (that is, the muon surface density at ground) \footnote{Another observable at ground is the angular distribution, which allows the reconstruction of the apparent production depth distribution by backtracking the muon trajectories to the shower axis. This will be studied in detail and will be published elsewhere \cite{a:MPDback}.}.  This knowledge is useful for fast air shower Monte Carlo simulations like CONEX \cite{Bergmann:2006yz}, and for use in the reconstruction algorithms of the number of muons, muon production depth, production energy and possibly the transverse momentum distributions, all of them directly inherited from the hadronic cascade.
The distributions of all muons at the moment of production, that is, including those that would decay later on flight, exhibits the most universal features independently of the observational conditions.

The inference of the fundamental distributions of all muons at production from the observed distributions it is not straightforward, since the information carried by  muons below some energy threshold - defined by the amount of matter transversed in the atmosphere - is completely lost. Nevertheless, by combining different observation conditions (zenith angle and distance to core) which correspond to  different effective energy thresholds, one might  constrain  the average distributions at production. Notice also that the propagation of muons itself does not depend on the details of the hadronic interactions models, being a completely decoupled problem.

A detailed understanding of the detector response to the different particles of an air shower is necessary to properly identify and interpret the muon information. Such a study is out of the scope of this paper.

This paper is organized as follows. In section 2 we define some quantities and describe the muon distributions at production. In section 3 we describe the propagation of muons and the approximations used. In section 4 we analyze the distributions at ground after propagation and compare them to a full air shower Monte Carlo simulation. In section 5 we discuss the effects of averaging the energy and transverse momentum distributions. Finally, we comment on the prospects and conclusions.

\section{The production of muons in EAS}

\begin{figure}[!h]
  \begin{center}
    \includegraphics[width=10cm]{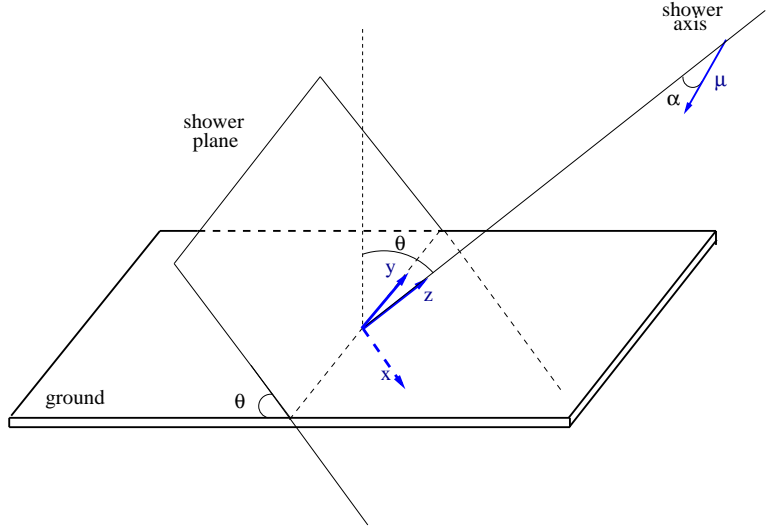}
    \caption[]{Scheme showing the shower plane (perpendicular to the shower axis), the ground surface and the system of coordinates.}
    \label{f:MPD_coord_3_final}
  \end{center}
\end{figure}

When a cosmic ray enters the atmosphere it creates an air shower of particles. The extrapolation of the original trajectory defines the so called {\it shower axis} and its intersection with the ground surface defines the {\it shower core}. We will use a Cartesian coordinate system  which is centered at the core position in ground, with the $z$-axis parallel to the shower axis. The $y$ axis will be parallel to ground, and the $x$-axis is positive downwards, entering the earth with an angle $\theta$ with the surface, which is also the angle between the $z$-axis and the zenith's direction (see Fig. \ref{f:MPD_coord_3_final}). A cylindrical coordinate system can be defined by $r=\sqrt{x^2+y^2}$ and $\zeta=\arctan{y/x}$ and it will be sometimes used for convenience.

The $z$-coordinate can be expressed in terms of the amount of matter along the shower axis from the top of the atmosphere:
\beq
X=\int^\infty_z \rho(z') dz'.
\eeq
We will use indistinctly $z$ and $X$ to express the position along the shower axis where muons are produced. The $z$ variable is more suited for calculations regarding geometry or kinematics, whereas the $X$ variable is used for the evolution of the cascade.

\begin{figure}[h]
\begin{center}$
\begin{array}{cc}
\includegraphics[width=7cm]{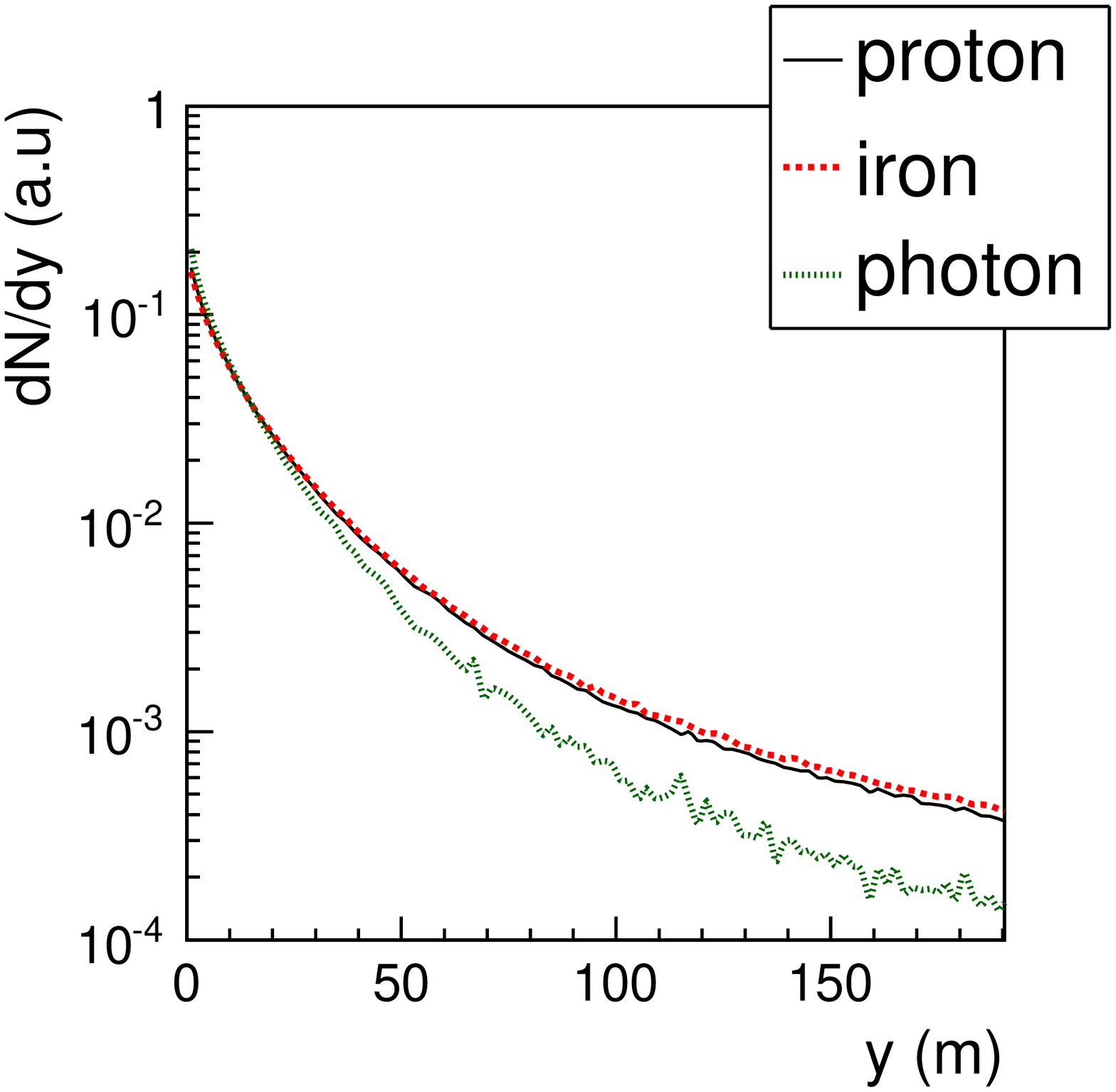} &
\includegraphics[width=7cm]{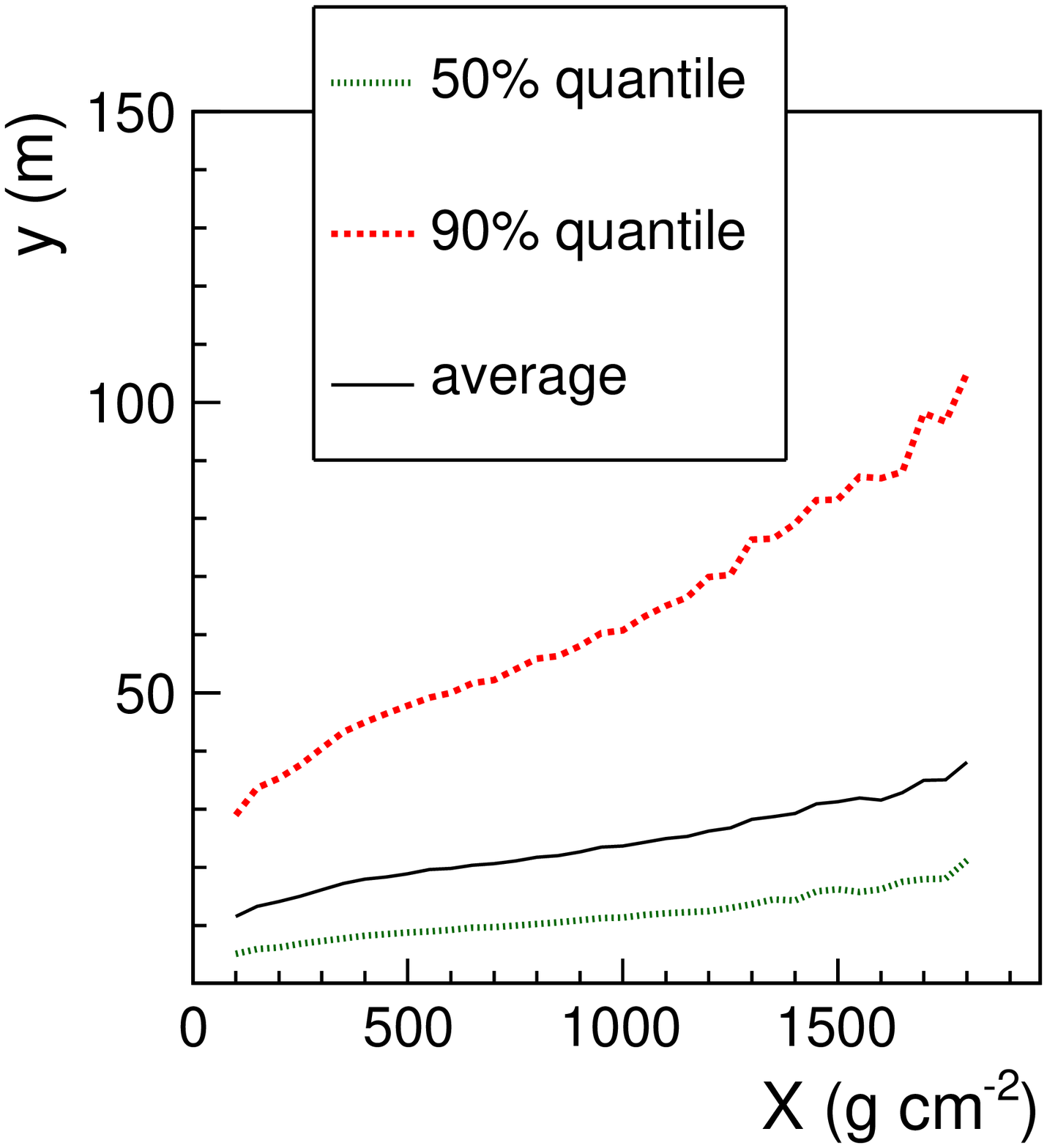}
\end{array}$
\end{center}
\caption{Left Panel: average distribution of the positive $y$-coordinate at the production point of muons for different primaries at $10^{19}$ eV at 60 deg. Right panel: average, median and 90\% quantiles of the $y$ distribution for different depths for a proton shower of $10^{19}$ eV at 60 deg.} 
\label{f:Compare_Depths_and_p_Fe_g_dNdyXUSP}
\end{figure}

In \cite{a:TimeModel}, it was argued that the transverse position of the production of muons, thus of the parent mesons decay, is confined to a relatively narrow cylinder. Fig. \ref{f:Compare_Depths_and_p_Fe_g_dNdyXUSP} right panel, displays the distribution of the y-coordinate  where muons were produced (the shower axis is at y=0) for different primaries. On the right panel, the average and the $y$-coordinate containing 50\% and 90\% of the production points are displayed as a function of the atmospheric depth.   The average value is of tens of meters. This distance is small when compared to the distances involved in EAS experiments, which span from hundreds of meters to several kilometers in the perpendicular plane. For instance, the Pierre Auger Observatory has its tanks separated by 1.5 km. Therefore, the position where the muon has been produced can be approximated by $(0,0,z)$, or simply $z$.
\begin{figure}[!h]
  \begin{center}$
\begin{array}{cc}
    \includegraphics[width=7cm]{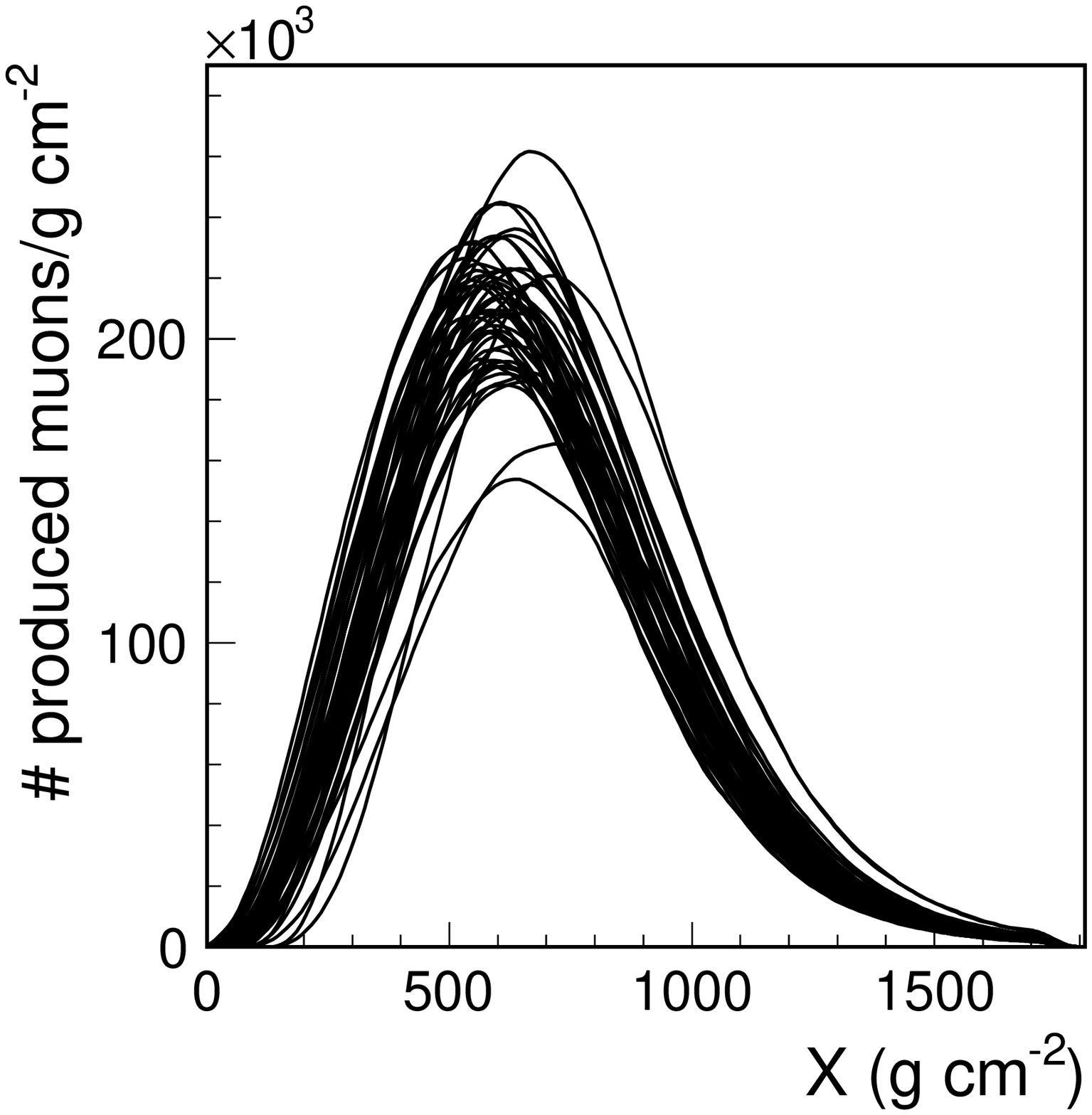}&
    \includegraphics[width=7cm]{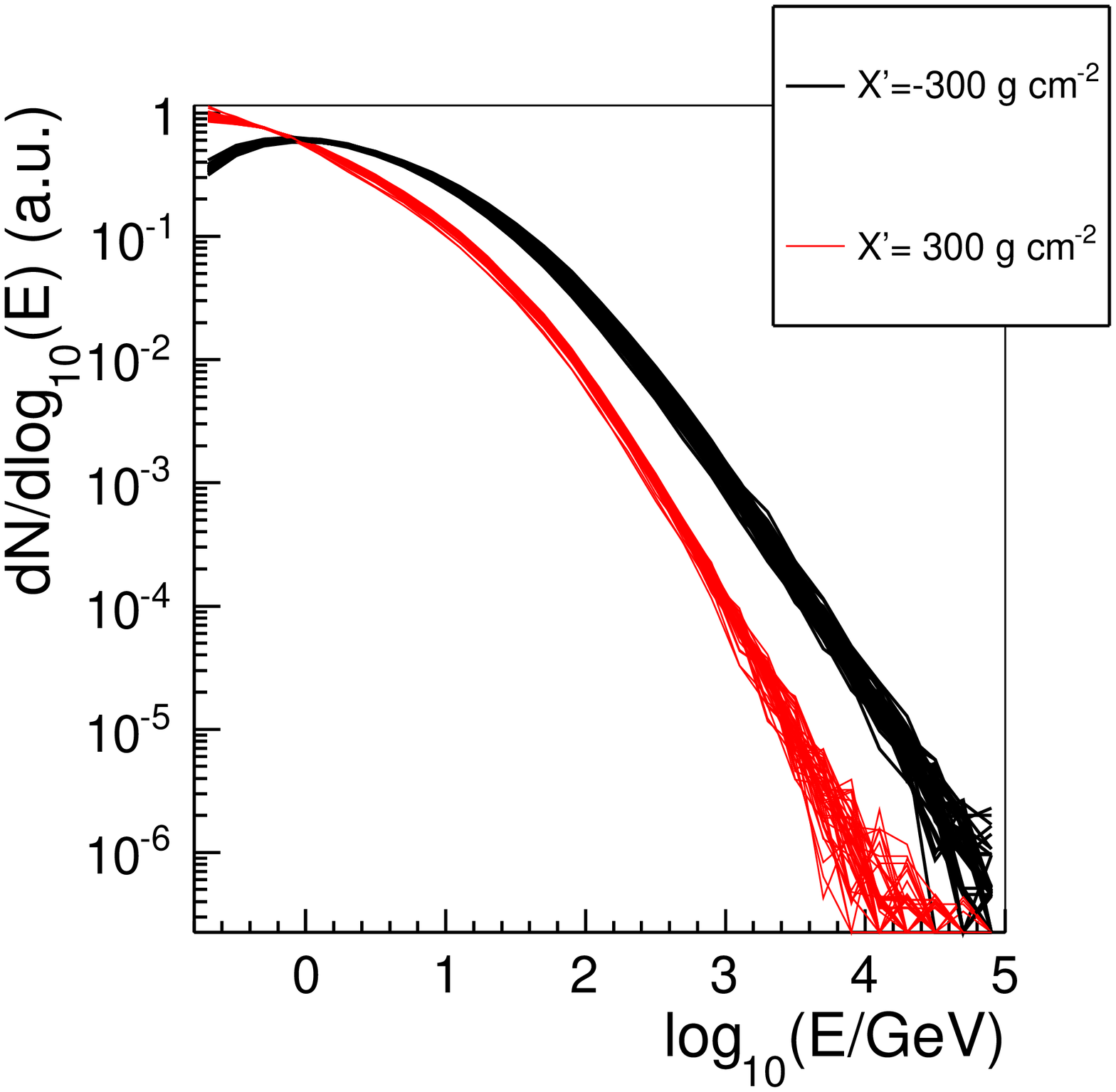}
\end{array}$
    \caption[]{Left panel:total number of muons produced per $\gcm$ ($h(X)$) for 50 proton showers at $10^{19}$ eV and 60 deg. Right Panel: normalized spectrum of muons at production at two distances to the shower maximum, $X'=X-\Xmax=-300$ $\gcm$ and $X'=X-\Xmax=300$ $\gcm$, for the same showers.}
    \label{f:hX_and_spectrum}
  \end{center}
\end{figure}

Every $dX$ along the shower axis, $dN$ muons are produced within a given energy and transverse momentum interval $dE_i$ and $dp_t$. Their overall distribution at production can be described in general with a 3-dimensional function, as:
\beq
\frac{d^{3} N}{dX\, dE_i\,dcp_{t}}= F(X,E_i,cp_{t})
\label{eq:totaldistribution}
\eeq
These 3D-distributions were recorded during simulation with CORSIKA v6.980 \cite{CORSIKA} at the moment of production, along with the standard ground particle output files. These distributions define most of the knowledge about muons contained in the air showers.

A library of CORSIKA v6.980 showers was created, with samples of 50 proton showers at 0, 40, 60 and 70 degrees with \QII \, \cite{Ostapchenko:2005nj,Ostapchenko:2006vr} model and energy $10^{19}$ eV. The relative thinning was set to $10^{-6}$, the maximum weight was set to $10^4$  and the hadronic maximum weight of $10^2$. The inner radial thinning was set to 1 cm. The kinetic energy cuts where set to 0.05 GeV for muons and hadrons and 0.003 GeV for electrons and photons. The altitude was set to 1400 m a.s.l, with the magnetic field of Malargue, Argentina.

 At 60 degrees, we also run 50 shower samples of iron and photon primaries with \QII, and for iron and proton for the models \SYB \, \cite{Ahn:2009wx} and \EPOS \, \cite{Werner:2005jf,Pierog:2006qv}. We also run proton showers at $10^{18.5}$ eV, $10^{19}$ eV and $10^{19.5}$ eV with \QII. 

A sub-sample of proton \QII \, $10^{19}$ eV showers was simulated with a lower maximum weight of $10^3$ for EM particles and 10 for hadrons (and muons). This subset was used to check the effects of the thinning, which are negligible.

The projection into the $X$ (or $z$) axis becomes
\beq
h(X)=\int F(X,E_i,cp_{t}) dE_i dcp_t
\eeq
and it is the so called {\it total/true} Muon Production Depth (Distance) distribution, or MPD-distribution for short.  It does not depend on the observational conditions since it does not contain any propagation effects of muons through the atmosphere. 
 Notice that this is different from the MPD-distributions of detected muons at a given position on ground  $\frac{dN}{dX}|_{(r,\zeta)}$, which includes the effects of propagation, as it will be explained later. This distribution is sometimes referred to as {\it apparent} MPD-distribution.

The total number of muons produced in a shower is
\beq
{\cal N}_0=\int h(X) dX
\eeq
It should be noted that this number is intrinsically different from the surviving muons, which are affected by the fluctuations of the depth of the first interaction, which changes the distance traveled by muons to ground. Some of the techniques used by Auger \cite{Schmidt:2007vq} use a fixed distance to the shower core, so they can also be affected by the lateral spread of the parent mesons. Notice that in CORSIKA, the function $F(X,E_i,cp_t)$ is only known above a certain energy threshold, $E_i>E_{th}$. Therefore the  {\it total/true} MPD-distribution depends on $E_{th}$, becoming $h_{E_{th}}(X)$.  In the same manner, the total number of produced muons corresponds to the total number of muons above the energy threshold, ${\cal N}_{E_{th}}$.  The simulations of the present work have a total energy threshold  $E_{th}=0.155$ GeV, the lowest allowed by CORSIKA code. The differences in the distributions at observation level induced by this particular threshold are negligible because such low energy muons decay on flight before reaching ground. $E_{th}$ value must be specified when referring to ${\cal N}_{E_{th}}$ or $h_{E_{th}}$  for which the low energy muons can contribute in large amounts\cite{a:hXRuben}. Changing from a concrete value of $E_{th}$ into another is trivial provided that we know $F(X,E_i,cp_t)$.

Eq. \ref{eq:totaldistribution} can be factorized and expressed as the product
\beq
F(X,E_i,cp_{t})= h(X) \,f_X(E_i,cp_t)
\label{eq:fX}
\eeq
where the function $f_X(E_i,cp_t)=\frac{F(X,E_i,cp_{t})}{h(X)}$ becomes the normalized $E_i$ and $cp_t$ distribution at a given production depth $X$.

In the approximations made on \cite{a:TimeModel,a:MPD,a:PhDCazon}, $f_X$ did not depend on $X$ and it was factorized in 2 independent distributions on $E_i$ and $cp_{t}$. This allowed analytical approximations of the distributions at ground. In this work we have included these correlations, improving the accuracy of the energy, production depth, and time distributions at ground, and allowing for a proper description of the muon lateral distribution at ground.

The function $h(X)$ tracks the longitudinal development of the hadronic cascade and represents the production rate of muons per $\gcm$. Its shape and features are extensively discussed in \cite{a:hXRuben}. The depth at which $h(X)$ reaches a maximum is denoted as $\Xmax$. 
 $\Xmax$ correlates with the first interaction point $X_1$ which corresponds to the first interaction of the primary in the atmosphere and the start of the cascading process \cite{a:hXRuben}. The most important source of fluctuations in air showers corresponds to the fluctuations of $X_1$, which causes an overall displacement of the whole cascade at first approximation. The amount $X'=X-\Xmax$  defines the amount of matter with respect to the shower maximum. The distributions can be expressed in terms of $X'$, where the most important source of fluctuations has been eliminated, and only the remaining effects are present.

In Fig. \ref{f:hX_and_spectrum} (left panel) $h(X)$ is shown for a sample of 50 showers. The fluctuations on the normalization and on $\Xmax$ are clearly observed. In the right panel we can see the normalized energy spectrum for two values of $X'$, namely -300 $\gcm$ and 300 $\gcm$. Both the energy and the transverse momentum show similar features when referred to the same distance to the shower maximum, $X'$. From now on, whenever we average distributions at production we do it on $X'$, that is, matching the maxima.

\begin{figure}[!h]
  \begin{center}$
\begin{array}{cc}
    \includegraphics[width=7cm]{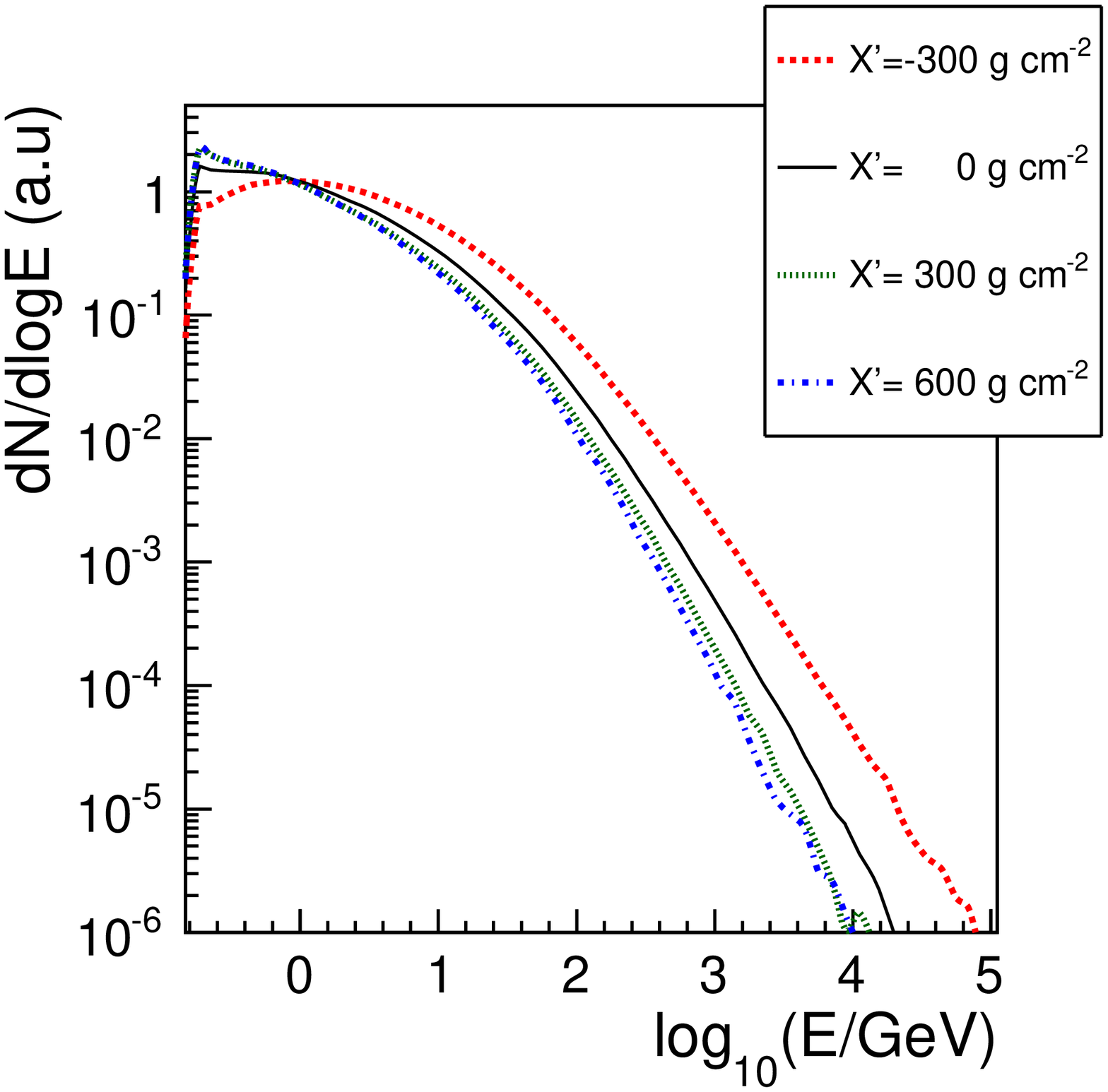}&
    \includegraphics[width=7cm]{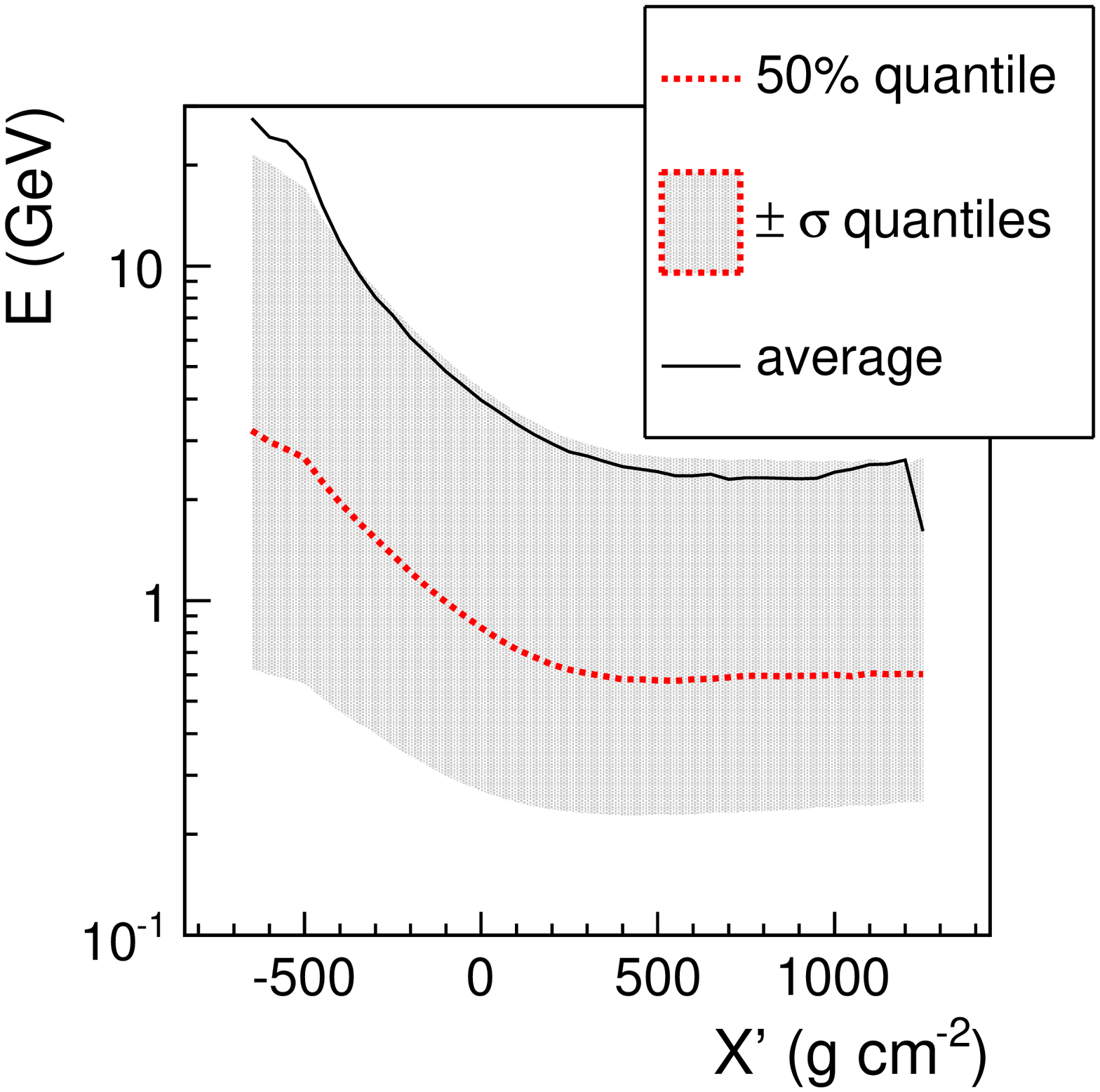}

\end{array}$
    \caption[]{Normalized average energy distribution of all muons at production for proton showers at $10^{19}$ eV and 60 deg zenith angle simulated with \QII. Left panel shows cuts at different $X'$ layers. Right panel:  50\% quantile (median) as a function of $X'$ along with a band defined by the 16\% and 84\% quantiles. The average value is also plotted as a solid  line.}
    \label{f:E_angles_Depths}
  \end{center}
\end{figure}

\begin{figure}[!h]
  \begin{center}$
\begin{array}{cc}
    \includegraphics[width=7cm]{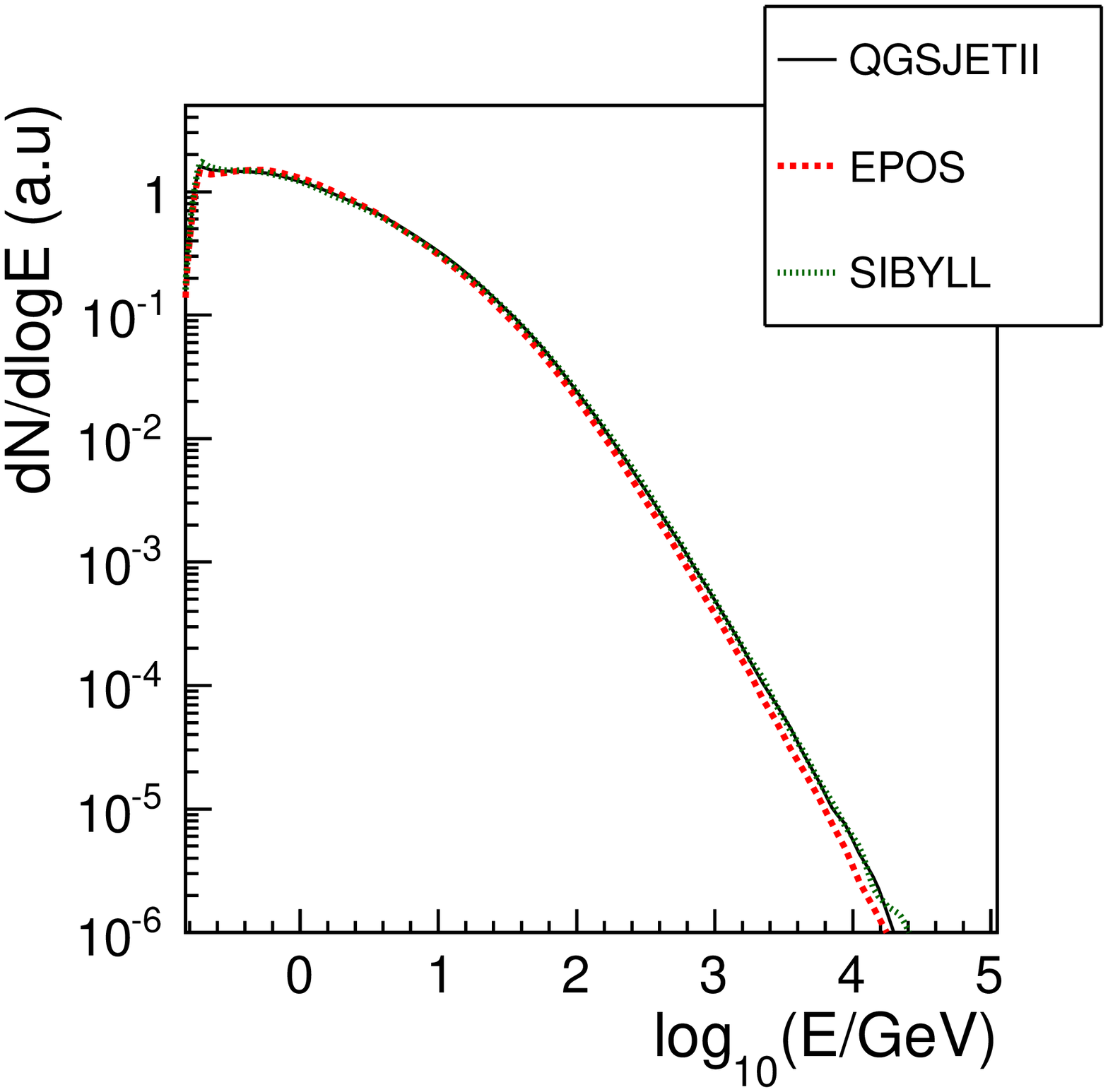}&
    \includegraphics[width=7cm]{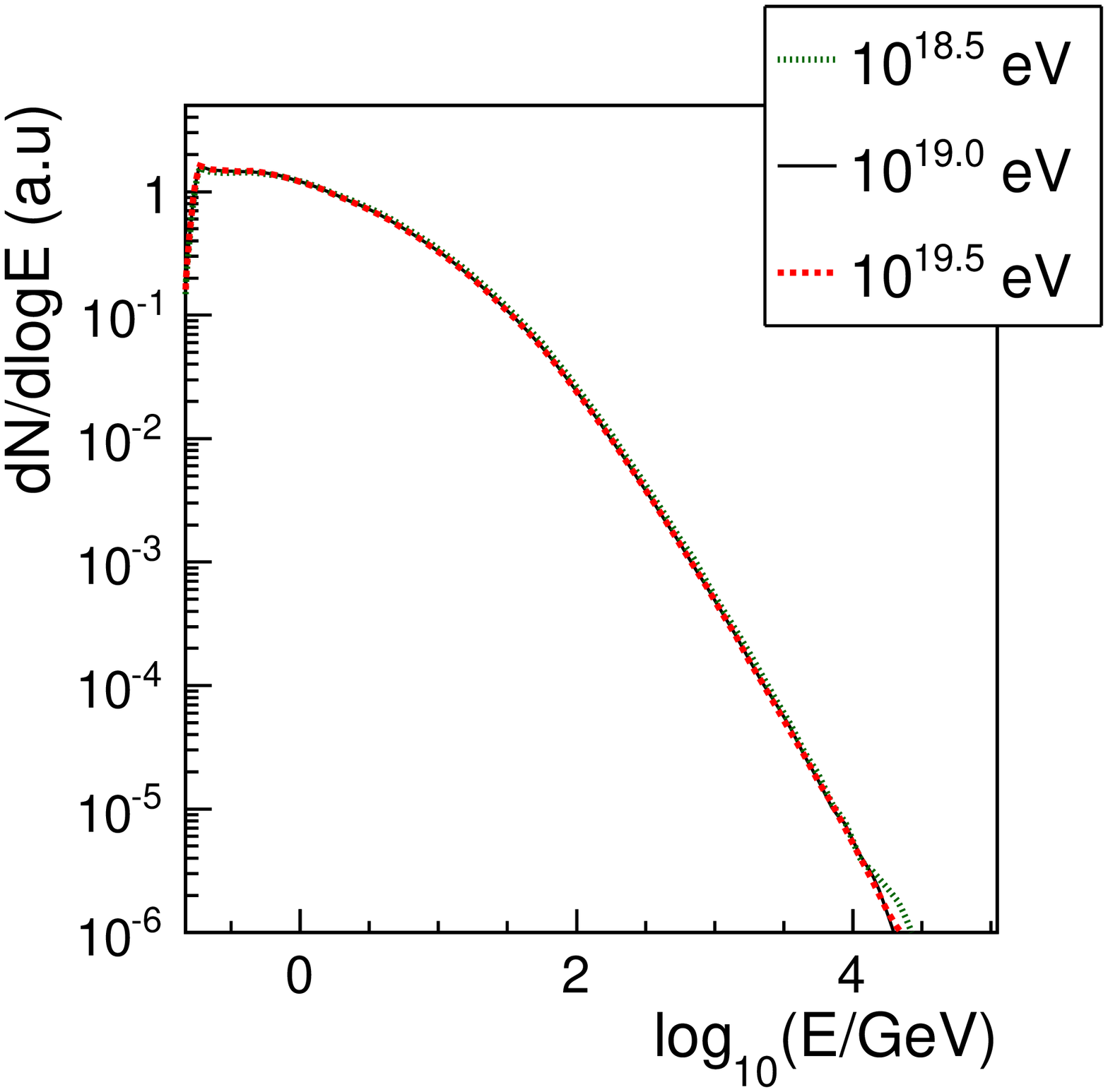}
\end{array}$
    \caption[]{Normalized average energy distribution of all muons at production for proton showers  at $X'=0$ $\gcm$ and 60 deg zenith angle. Left panel,  $10^{19}$ eV  for different hadronic interactions models. Right panel, \QII \, for different primary energies.}
    \label{f:E_Models}
  \end{center}
\end{figure}

In  \cite{a:TimeModel} and \cite{a:PhDCazon} the muon spectrum at production was approximated by a power law, $E_i^{-2.6}$, following the high energy tails of the pion production on the hadronic reactions. In that approach we did not have access to all muons at production, and the energy spectrum was extrapolated down to low energies with the same power law. In Fig. \ref{f:E_angles_Depths}, left panel, the actual average energy spectrum of all muons at production is displayed for proton showers at $10^{19}$ eV in different $X'$ layers. At low energies the single power law clearly does not work. In addition, the energy spectrum evolves with $X'$ by becoming softer, and stabilizing the shape after the shower maximum. 
\begin{figure}[!h]
  \begin{center}$
\begin{array}{cc}
    
    \includegraphics[width=7cm]{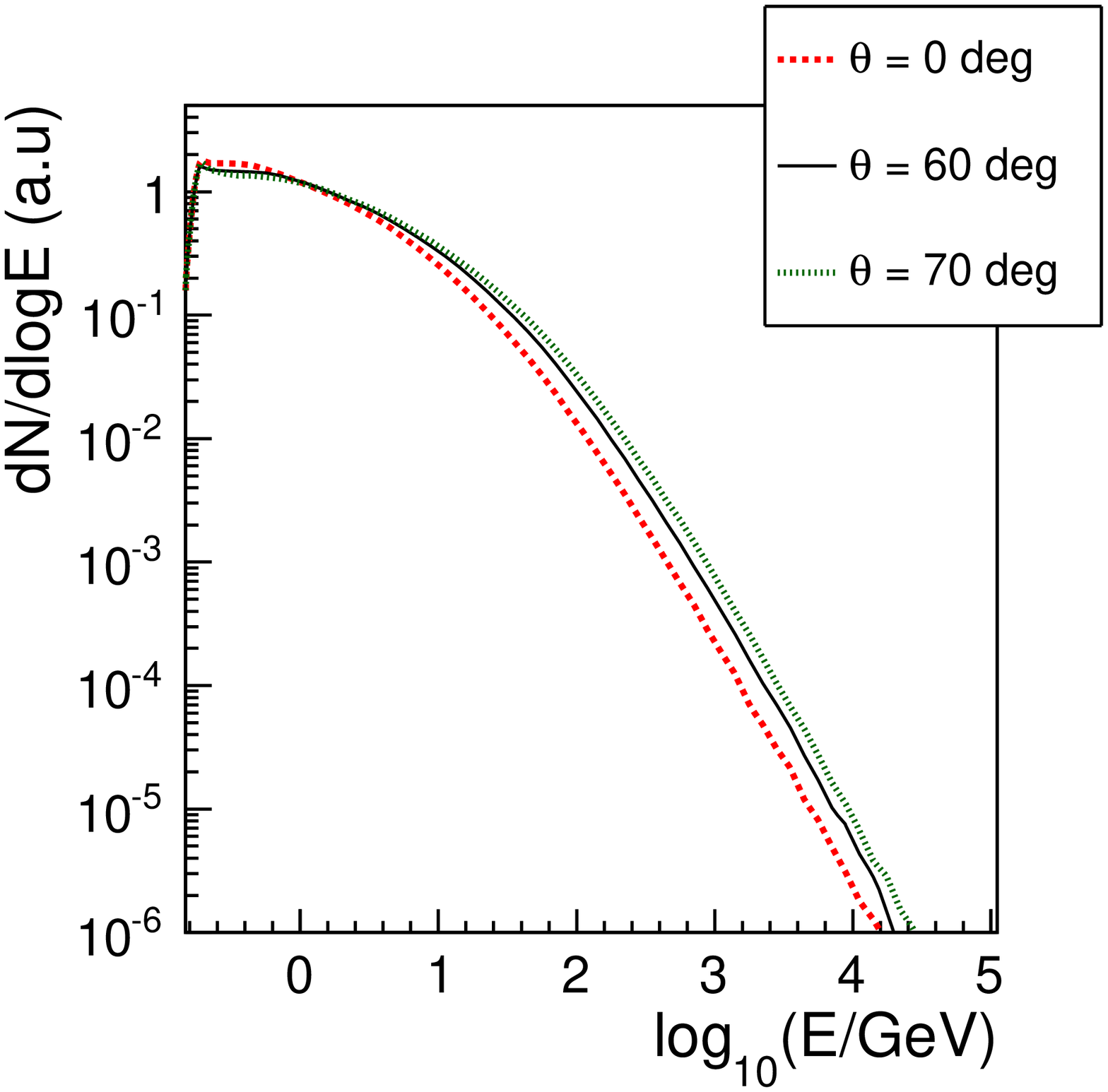}&
\includegraphics[width=7cm]{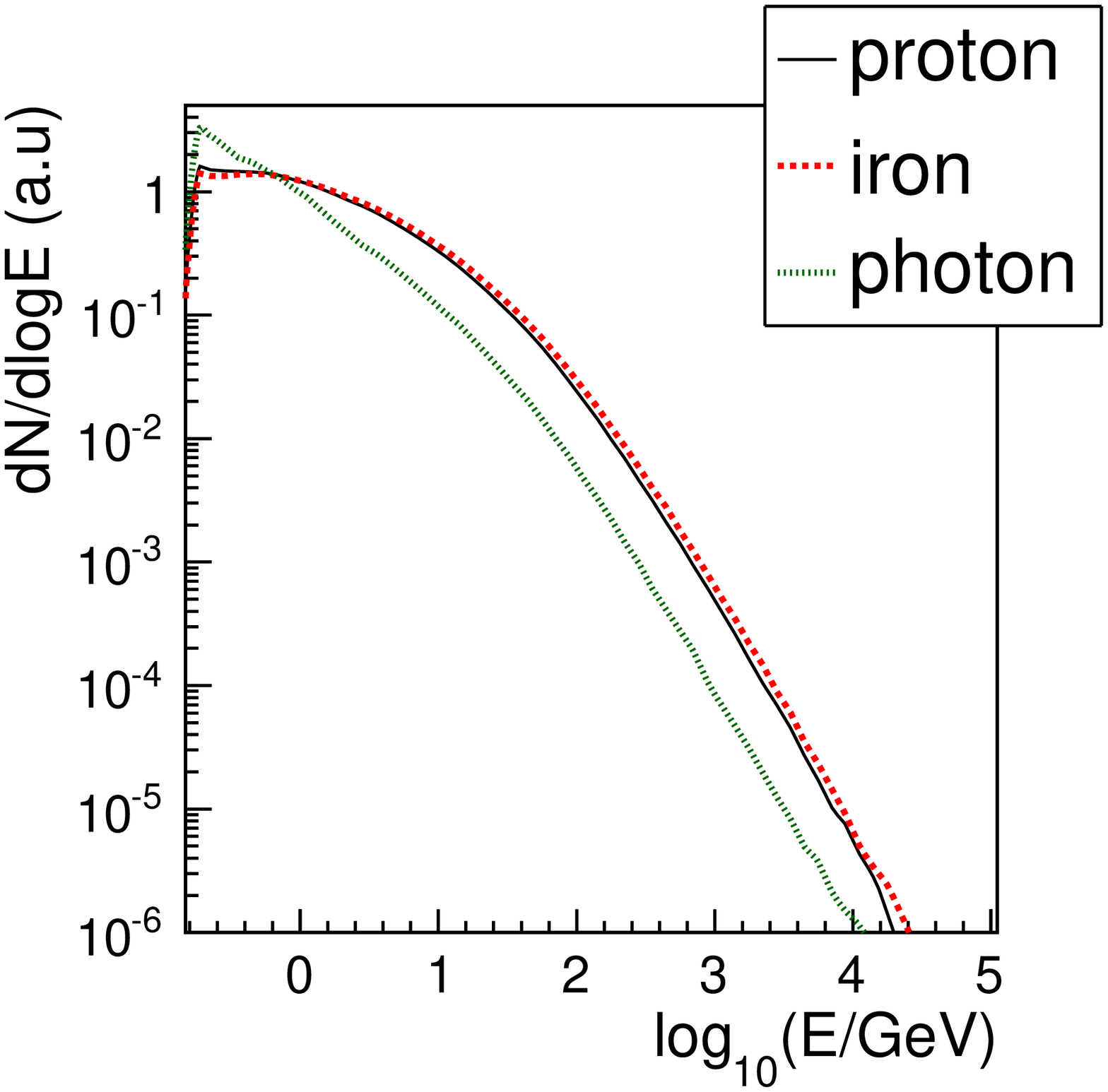}
\end{array}$
    \caption[]{Normalized average energy distribution of all muons at production for proton initiated showers at $10^{19}$ eV simulated with \QII \. for different zenith angles (left panel), and different primaries at 60 deg zenith angle (right panel).}
    \label{f:E_primaries_angles}
  \end{center}
\end{figure}
In Fig. \ref{f:E_primaries_angles}, left panel, the energy spectrum is displayed for different zenith angles at $X'=0$, showing a mild dependence, becoming harder at higher zenith angles. This might be due to the higher critical energy of the pions at higher zenith angles, given that the shower develops in less dense air in average. Regarding the dependence with the primaries, Fig. \ref{f:E_primaries_angles}, right panel displays the spectrum for proton, iron and photon primaries at $X'=0$. Whereas proton and iron curves practically overlap, photons, on the other hand, produce a much softer energy spectrum. The mechanisms for production of muons in photon showers is basically photo-pion production, so they are intimately related to the energy spectrum of the EM cascade. Lastly, Fig. \ref{f:E_Models}, left panel, displays a comparison between different models, in the same conditions. While \QII\, and \SYB \, overlap, \EPOS \, shows a slightly different behavior, with a softer spectrum.  Fig. \ref{f:E_Models}, right panel, displays a comparison of different primary energies with the same hadronic interaction model, where the curves practically overlap.

\begin{figure}[!h]
  \begin{center}$
\begin{array}{cc}
    \includegraphics[width=7cm]{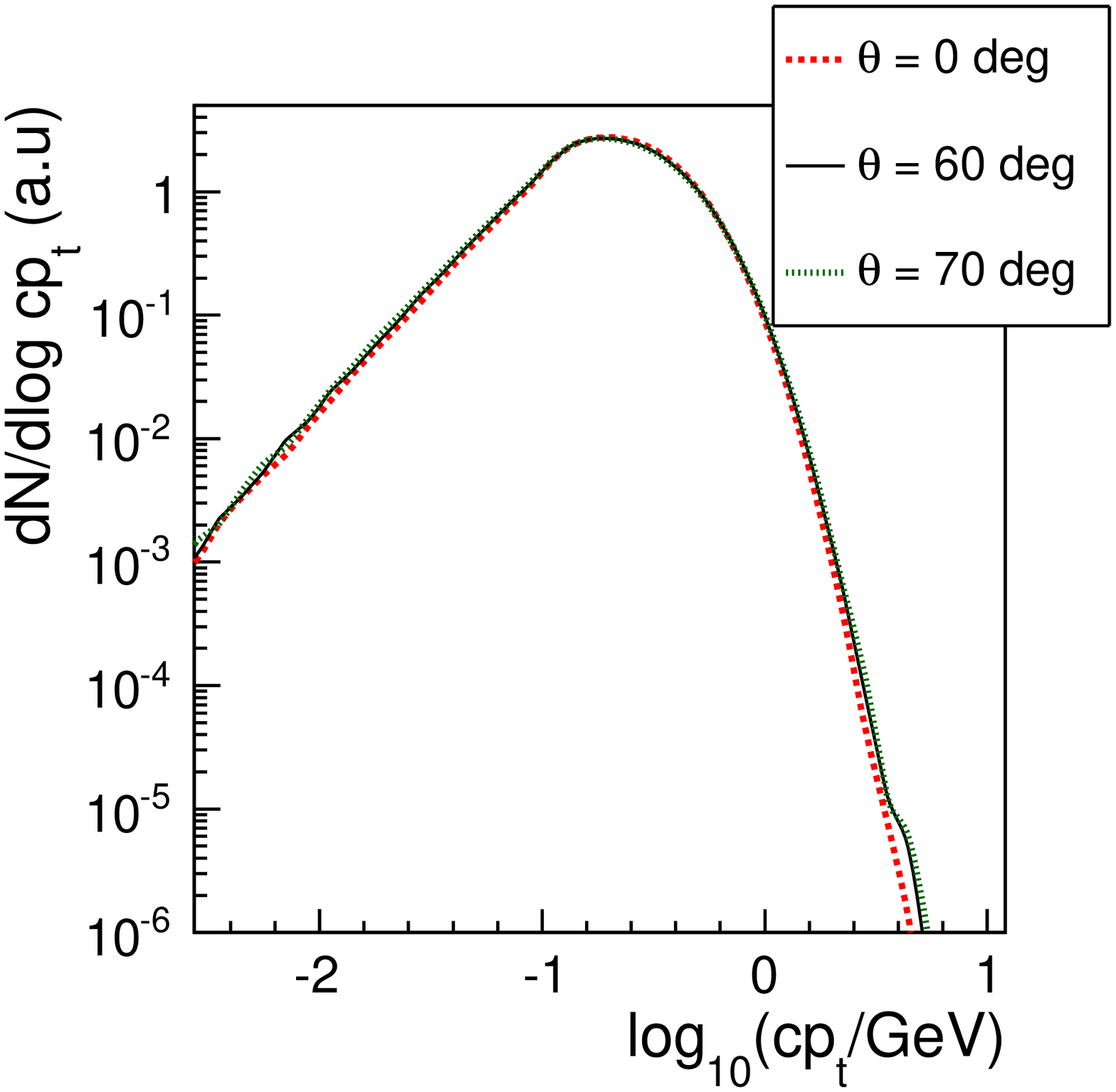}&
    \includegraphics[width=7cm]{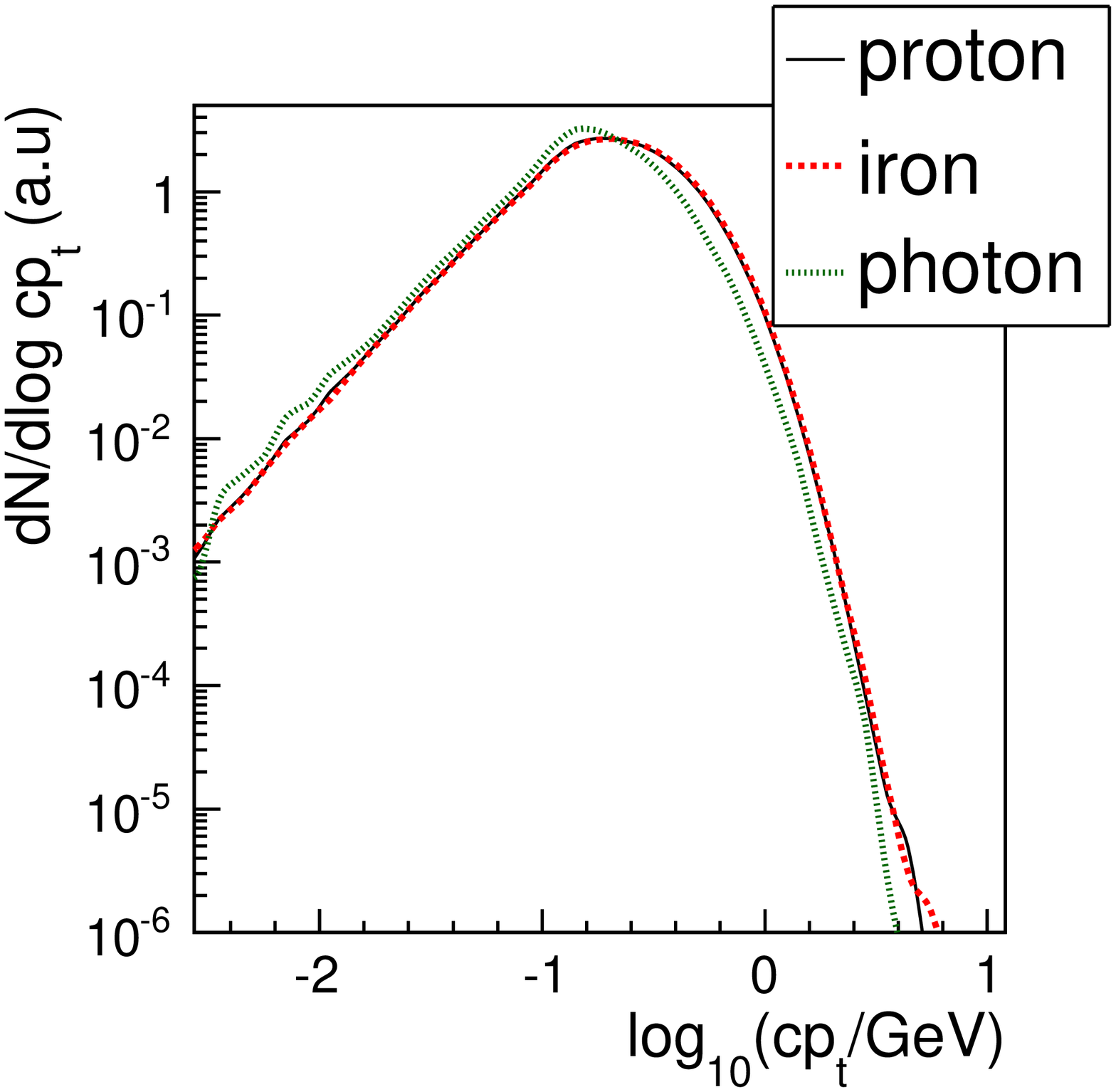}
\end{array}$
    \caption[]{Normalized average $p_t$ distribution of all muons at production for showers at $10^{19}$ eV at $X'=0$ $\gcm$ simulated with \QII. Left panel, different zenith angles, with proton primaries. Right panel, different primaries at 60 deg zenith angle.}
    \label{f:pt_degs_primaries}
  \end{center}
\end{figure}

\begin{figure}[!h]
  \begin{center}$
\begin{array}{cc}

 \includegraphics[width=7cm]{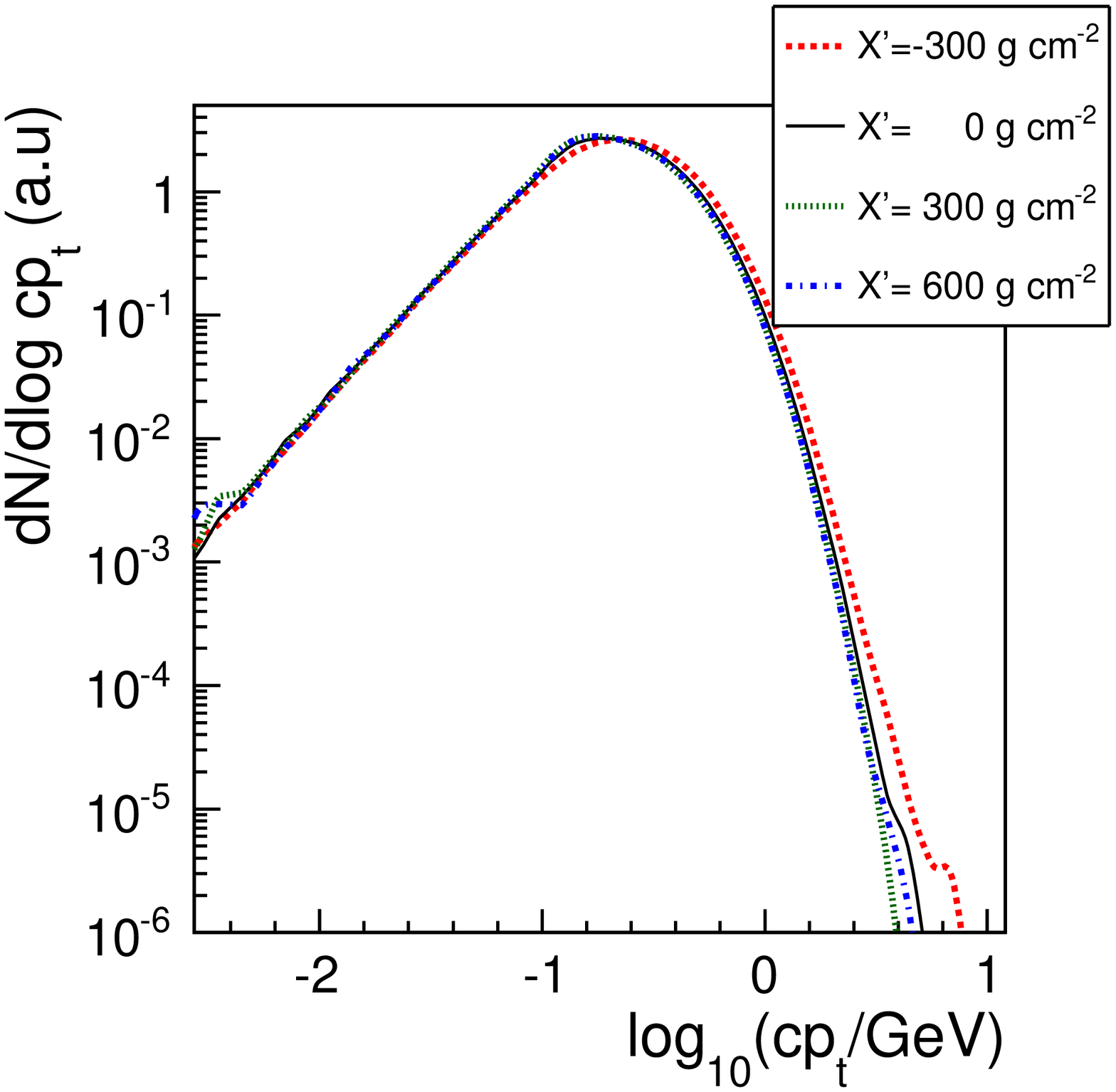}&
    \includegraphics[width=7cm]{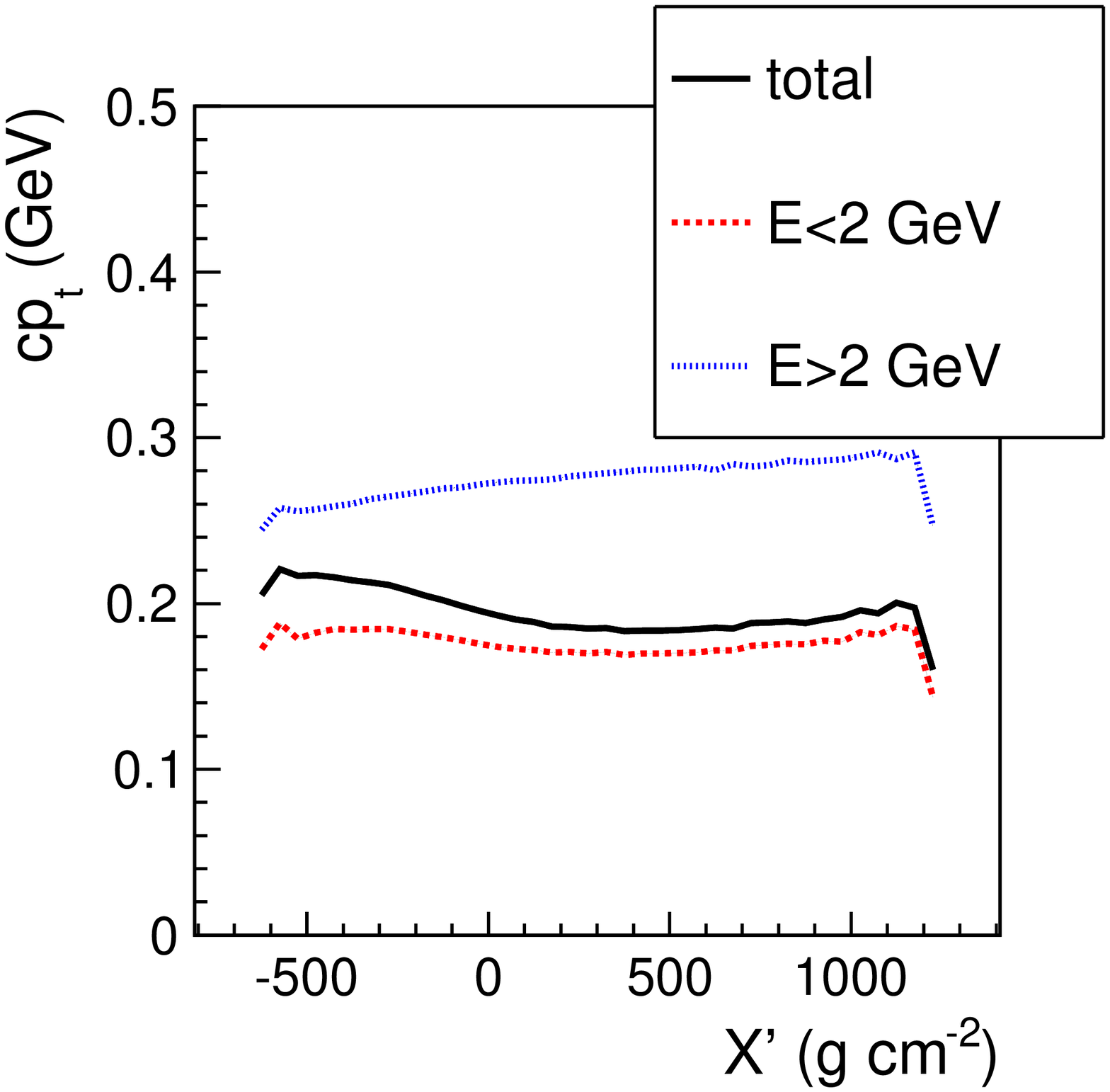}

\end{array}$
    \caption[]{Normalized average $p_t$ distribution of all muons at production for proton initiated showers at $10^{19}$ eV simulated with \QII \, for different values of $X'$ (left panel). Median $cp_t$ as a function of $X'$ for the total number of muons, and the median for those muons with $E_i>2$ GeV and $E_i<2$ GeV, as labeled (right panel).}
    \label{f:pt_depths_XEcuts}
  \end{center}
\end{figure}

The transverse momentum distributions are responsible for most of the lateral displacement of muons with respect to the shower axis. In \cite{a:MPD}, the $p_t$ distributions were approximated by an unique function,  $dN/dp_t=p_t/Q^2 \exp(-p_t/Q)$, independent of the energy of the muon and its production depth, primary mass and zenith angle. In the current work, we uncover in detail all the dependencies.
 The $p_t$ distributions display a quite universal shape as a function of the zenith angle, (see left panel of Fig. \ref{f:pt_degs_primaries} displaying the normalized average distributions of the $p_t$ distributions at $X'=0$ $\gcm$), and do not depend on the primary (right panel Fig. \ref{f:pt_degs_primaries}), except for the case of photons, which respond to a completely different muon production mechanism, as explained before. As the shower evolves, the $p_t$ spectrum becomes softer (Fig. \ref{f:pt_depths_XEcuts}, left panel shows the evolution as a function of $X'$).  Besides this dependence on $X'$, the $p_t$ distributions also depend on the energy of the muons. The right panel shows the medians of  the $p_t$-distribution for 2 different energy cuts. The low energy muons display a smaller $p_t$, and at high energies, the $p_t$ distribution prefers higher $p_t$ values. Both distributions show a different dependence on $X'$. We have found that the different correlations of the $p_t$ with $E_i$ and $X$ must be included into the model in order to properly predict the muon lateral distribution at ground.

\begin{figure}[!h]
  \begin{center}$
\begin{array}{cc}
    \includegraphics[width=7cm]{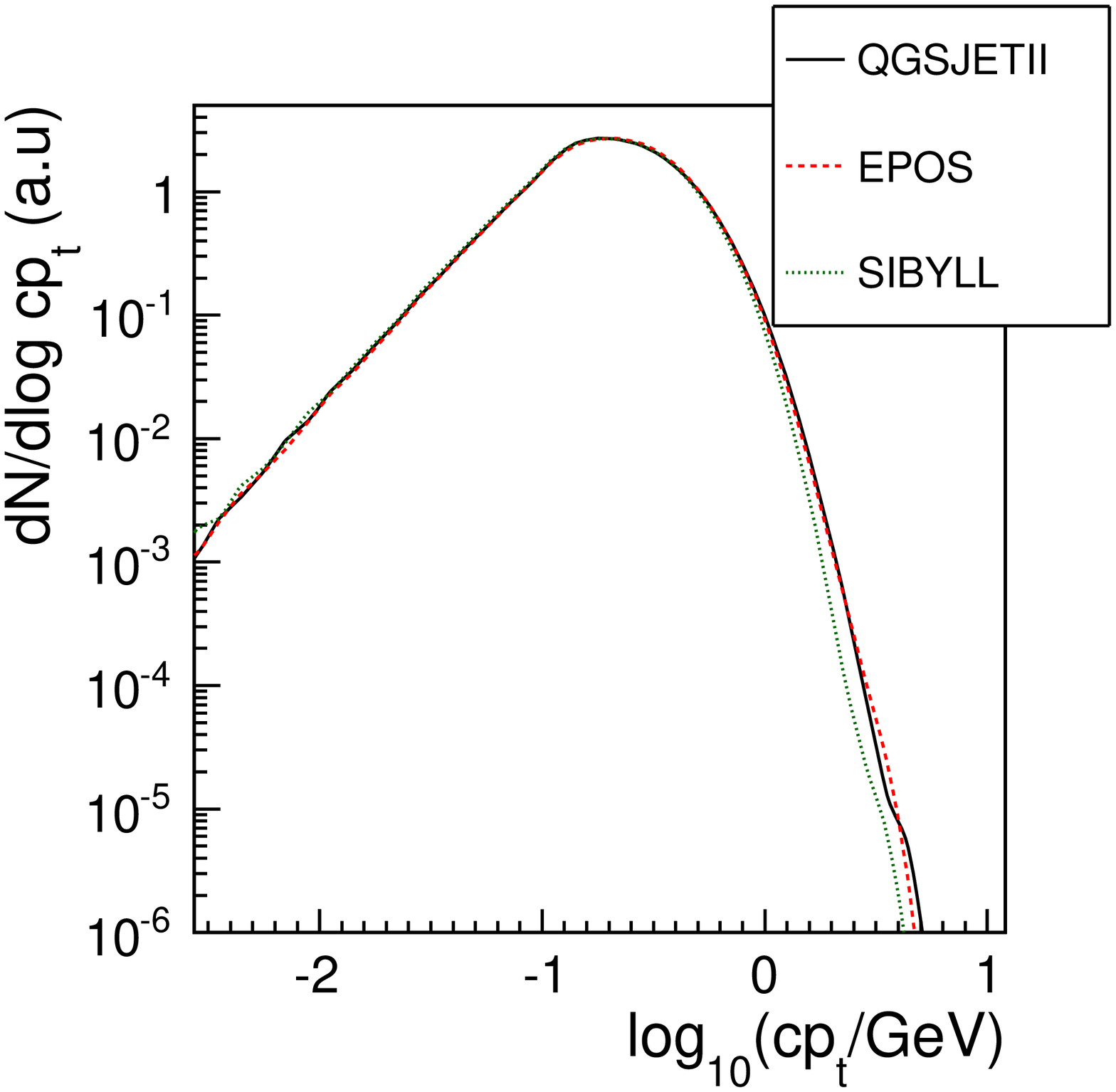}
    \includegraphics[width=7cm]{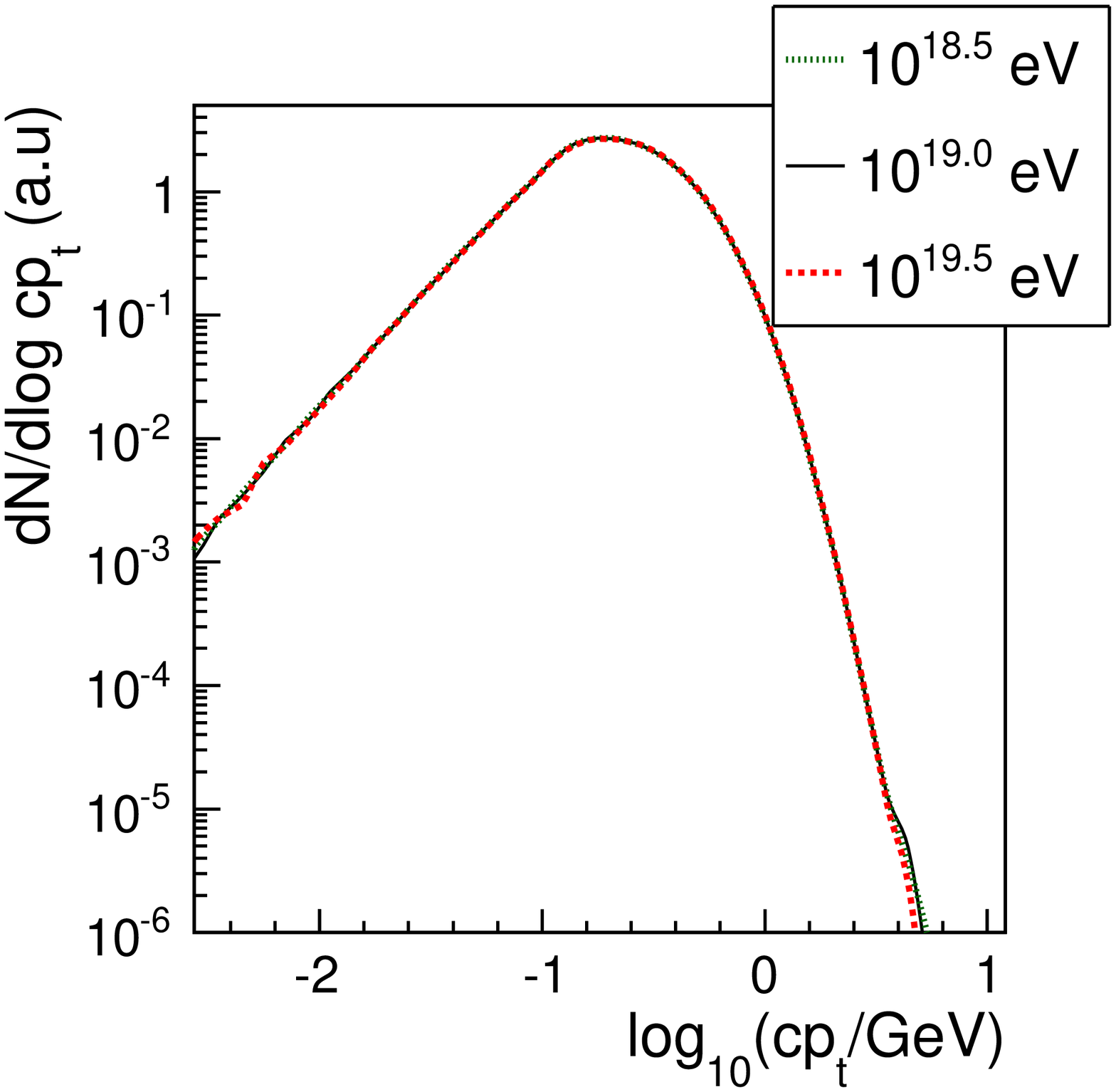}
\end{array}$
    \caption[]{Normalized average $p_t$ distribution of all muons at production for proton showers  at $X'=0$ $\gcm$ and 60 deg zenith angle. Left panel,  $10^{19}$ eV  for different hadronic interactions models. Right panel, \QII \, for different primary energies.}
    \label{f:pt_Models}
  \end{center}
  \end{figure}

Fig. \ref{f:pt_Models}, left panel, displays the $p_t$ distribution for 3 models. The high $p_t$ tail of \SYB \, is suppressed  with respect to the other models, which practically overlap.  Fig. \ref{f:pt_Models}, right panel, displays a comparison of different primary energies with the same hadronic interaction model, where the curves practically overlap.

\section{Propagation of muons through the atmosphere}

The  propagation of muons in matter below 1 TeV is reviewed in \cite{a:PDG}. In this section we will describe the aspects relevant for the distributions of muons at ground in UHECR-induced air showers. The extremely energetic muons coming from the first interactions would need a huge area covered by high resolution detectors in order to collect and identify them. On the contrary, the region below 1 TeV constitutes the bulk of the muons, and it is the most important from the statistics point of view.
 
The transport of muons to ground was implemented by means of a fast Monte Carlo in two steps. In the first step the muons were propagated to ground following a straight line according to their 3-momentum and the continuous energy loss was calculated. In the second step, the multiple scattering and magnetic field effects were included, the impact point on ground was corrected, and the energy loss  reevaluated. Then, the probability of decay and the time delay for the corrected energy loss and trajectory were obtained.

 For each shower, the variables at the production point, namely  $z$, $E_i$, and $cp_t$ were sampled from the distributions $F(z,E_i,cp_t)$ with a weight $w_i$. The momentum of the muon defines an angle $\alpha$ respect to  the  shower axis given by
\beq
\sin\alpha \simeq\frac{c p_t}{E_i},
\label{eq:geometry_relation}
\eeq
where we took the approximation $cp\simeq E_i$. A 10 GeV (1 GeV) muon typically will span a 1 deg (10 deg) outgoing angle with respect to the shower axis. The outgoing polar angle follows a symmetric distribution $\frac{dN}{d\zeta}=\frac{1}{2 \pi}$ that was also randomly sampled.

From the production point $(0,0,z)$ a straight line trajectory defined by $(\alpha,\zeta)$ is extrapolated until it hits ground, defining the arrival position of the muon $(r,\zeta,\Delta)$, where $\Delta$ is the $z$ coordinate of the ground in the shower system, and $r$ is the distance to the core in the perpendicular plane. For a flat ground surface they  become
\beq
r=\frac{z}{\cos\zeta \tan\theta+\frac{1}{\tan\alpha}}
\eeq
\beq
\Delta=r \cos \zeta \tan \theta
\eeq
which can be generalized for a curved earth surface.
The distance traveled by the muon from the production point to the ground is
\beq
l=\sqrt{r^2+(z-\Delta)^2}
\eeq

As stated in \cite{a:TimeModel}, we can define a plane parallel to the $xy$ plane that travels at the speed of light that contained the first interaction point and hits ground at $t=0$. Assuming that all particles travel at speed of light and no delay is accumulated during the hadronic cascade, the muon is produced with no delay. The arrival time delay of the muon respect to this plane front becomes
\beq
ct_g=l-(z-\Delta)
\label{eq:tg}
\eeq
which can be approximated by $ct_g \simeq \frac{1}{2}\frac{r^2}{z-\Delta}$ in most practical cases.
A correction due to the path traveled by the parent meson was applied. The trajectory would have started an amount $\Delta z_{\pi}=c \tau_{\pi} \frac{E_i}{m_{\pi}c^2} \cos \alpha$ higher up in the atmosphere. Thus, we replace $z$ (and $l$) in Eq. \ref{eq:tg} by $z+\Delta z_{\pi}$.  A muon produced at $z=10$ km from ground and reaching $r=1000$ m will be delayed $t_g\simeq$ 165 ns because of geometric effects, ($t_g\simeq$ 167 ns if we did not take the parent meson correction).

Along this trajectory the muon will suffer a number of processes, namely: energy loss, multiple scattering, and magnetic field deflections. Each one of these processes will modify the momentum, trajectory and time delay with respect to the plane front traveling at velocity $c$. In addition, muons can decay on flight with a certain probability, in which case some information is lost.
\subsection{Continuous energy loss in the atmosphere}
Muons lose energy as they travel through the atmosphere \cite{a:PDG}. Only the energy loss by ionization is taken into account as:
\beq
\frac{dE}{dX}= -a
\eeq
and where $a$ has been parametrized as
\beq
a=\left(2.06+0.5453 \xi+\frac{0.0324}{(\xi+1.0312)^2}\right) \times 10^{-3} \,\,\,{\rm (GeV/ \gcm)}
\eeq
where $\xi=\log_{10}(E/{\rm GeV})$. For energies  $E> \sim 50$ GeV, where the radiative losses start to appear, the relative error on the energy loss to the total energy are neglected.

$X(z)$ can be calculated with the detailed description of the atmosphere in different layers currently used by most Monte Carlo codes. Nevertheless, we used a compact description of the atmosphere in order to calculate the energy losses of muons in a single step. We used the exponential approximation in a single layer, $\rho=\rho_0 e^{-\frac{h}{h_0}}$. The thickness of atmosphere traversed by a muon between the production point and ground becomes
\beq
\Delta X(l)=\int_0^l \rho(l) dl=\rho_0 l_0 \left(1-e^{-\frac{l}{l_0}} \right) 
\eeq
where $l=h/\cos \theta_\mu$ and $l_0=h_0/\cos\theta_\mu$. $h$ is the height and and $\theta_\mu$ is the zenith angle of the trajectory of the muon. The values of $\rho_0$ and $h_0$ were obtained by the best fit to $X(z)$ up to 10 km height. Notice that the region close to ground is the one that affects the most the propagation of muons.

 The energy of the muon evolves as
\beq
E(l)=E_i -a \rho_0 l_0 \left(1-e^{-\frac{l}{l_0}} \right)
\eeq

The finite energy of the muons induces a delay with respect to a particle traveling at the speed of light. The particles belonging to the hadronic shower decrease their energy on a geometric progression, due to the cascading process. The delay accumulated prior to the moment of decay into a muon is neglected. After the muon is produced, its kinematic delay becomes:
\beq
ct_\epsilon=\int_0^l \left[ 1-\left(\frac{mc^2}{E(l)}\right)\right]^{-\frac{1}{2}} dl' - l 
\eeq
 This equation can be easily integrated for the case where $\frac{mc^2}{E(l)}\ll 1$ yielding:
\beq
ct_\epsilon=\frac{1}{2} \frac{m^2c^4}{E_\infty} \left[-l_0\left(\frac{1}{E_i}-\frac{1}{E_f}\right)+\frac{l}{E_\infty}+\frac{l_0\log\frac{E_i}{E_f}}{E_\infty}\right]
\eeq
where 
\beq
E_f=E_i- a \rho_0 l_0 \left( 1- e^{-\frac{l}{l_0}}\right)
\eeq
and
\beq
E_\infty=E_f + \rho_0 a l_0
\eeq
Most of the delay of the particle is accumulated at the end of the trajectory, where the energy is lower due to 
the continuous energy loss. A 5 GeV (10 GeV) muon traveling 10 km with a zenith angle of 60 degrees, will reach ground with an energy $E_f=3.0$ GeV (7.8 GeV) and a kinematic delay of 12 ns (2.3 ns).

\subsection{Probability of Decay}
\label{s:decay}
Muons can decay on flight with a probability that depends on the energy. 
 The low energy muons are reduced or totally suppressed if they are below a certain energy threshold dependent on the amount of matter they would have to traverse to ground, which is of the order of 0.2 GeV every $\sim$ 100 $\gcm$. 
The probability of decay in a interval $dl$ is $- \frac{m c^2}{c \tau} \frac{1}{E} dl$.
By integrating, we obtain the probability of decay as a function of the traveled distance:
\beq
p(E_i,l)= \left(\frac{E_f}{E_i} \right)^{\frac{m c^2}{c \tau}\frac{l_0}{E_\infty}} e^{-\frac{m c^2}{c \tau}\frac{l}{E_\infty}}
\eeq
In our Monte Carlo, the weight of the muon at ground is substituted accordingly
\beq
w=w_i p(E_i,l)
\eeq
A 5 GeV (10 GeV) muon traveling 10 km to ground with a zenith angle of 60 degrees has a probability of survival $p=0.67\%$ (0.84\%).

Decay plays a fundamental role on shaping the distributions of muons at ground: the muon lateral distribution, time distribution and the energy spectrum are suppressed in the regions dominated by low energy muons, if we compared to the case where the decay was not present.

\subsection{Geomagnetic Field}
The magnetic field of the Earth affects the muons by bending their trajectory, changing the impact point on ground and delaying the arrival time. The effects are more visible the longer the trajectories.
 Muons travel long paths in inclined showers, reaching up to 220 km for showers at $86^{\circ}$ zenith angle. In \cite{Ave:2000xs} a model that
 describes this effect was developed. After neglecting the effects of the transverse momentum, $cp_t\ll E$ we obtain the radius of curvature $R$ of the trajectory of a muon of energy $E$:
\beq
 R =\frac{p}{e B_{\perp}}\simeq \frac{E}{c e B_{\perp}}
\label{eq:R_B}
\eeq
where $B_{\perp}$ is the projection of the magnetic field onto the perpendicular plane. The energy of the muon was approximated by $E=(E_i+E_f)/2$.

\begin{figure}[!htb]
\begin{center}
\includegraphics[width=10cm]{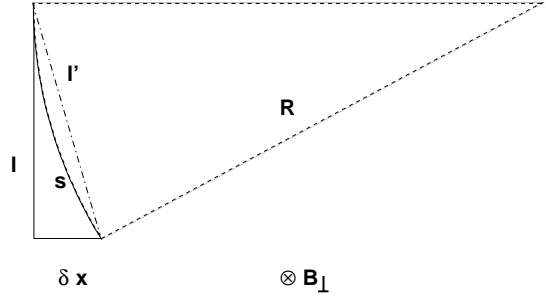}
\caption{Deviation from the expected trajectory of a positive particle
  (traveling downwards) because of the magnetic field  (entering the paper). The difference between the arc $s$ and a chord $l'$ introduces a time delay.  }
\label{f:B_deviation}
\end{center} 
\end{figure}

 Let us now define a set of coordinates $(x_B,y_B)$ in the shower plane in such a
way that the positive $y_B$ axis is aligned with $\vec{B}_\perp$. As a muon travels  it will be shifted $\delta x_B$ in the
direction perpendicular to $\vec{B}_\perp$ given by (see Fig.~\ref{f:B_deviation}):
\begin{equation}
\delta x_B= R\left[1-\sqrt{1 \mp \left(\frac{l}{R}\right)^2} \right]\simeq \pm \frac{1}{2} \frac{l^2}{R}
\end{equation}
 The $\pm$ sign depends
on the charge of the muon, which is sampled randomly with equal probability. This means that  the locations of arriving muons are shifted in dependence on the traveled distance $l$ but
 also on the energy of the muon. The coordinate $y_B$ is left unchanged. For a typical value of $B_{\perp}=10\,\, {\rm \mu T}$, a 5 GeV (10 GeV) muon that travels 10 km deviates $\delta x_B=83$ m (16.7 m).

The actual path length of the trajectory is enlarged as
\beq
s(l)=l+2 R  \arcsin \frac{l}{2 R}\simeq l+ \frac{1}{24}\frac{l^3}{R^2}
\eeq
 The difference between $s(l)$ and $l$ is the so called {\it geomagnetic time delay}:
\begin{equation}
ct_{B}  \simeq \frac{1}{24}\frac{l^3}{R^2}
\end{equation}
A 5 GeV (10 GeV) muon traveling 10 km, will delay $t_{B}\simeq$ 0.04 ns (0.01 ns).

\subsection{Multiple Scattering}

Muons traversing the air are deflected by small-angle scatters. Most of  these deflections are due to Coulomb scattering from nuclei. The many small interaction add up to an angular and space deviation that follows a Gaussian distribution to a good approximation. 
\begin{figure}[!h]
  \begin{center}
         \includegraphics[width=7cm]{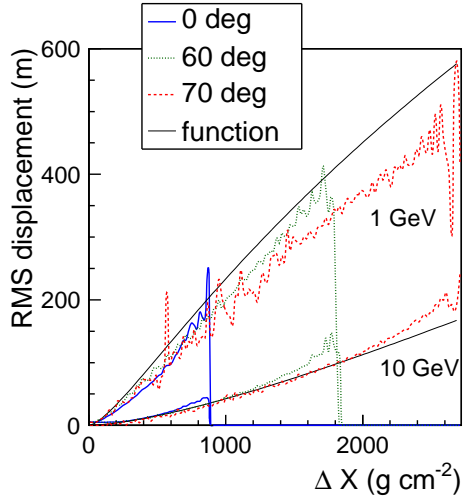}
	     \caption[]{Examples of RMS displacement due to multiple scattering for different zenith angles and $E_f=1$ GeV and $E_f=10$ GeV as labeled. The solid black line is the parametrization given in \cite{a:PDG} with $E=(E_i+E_f)/2$}
    \label{f:RMSy}
  \end{center}
\end{figure}

Using a dedicated library of CORSIKA simulations with the geomagnetic field off, we have created look-up tables (in the form of collections of 3D histogram) with  the different relevant scattering quantities, namely, the RMS of the lateral displacement, the RMS and mean of the logarithm of the time delays, and the RMS and mean scattering angle as a function of the energy at ground and total distance traveled by the muon. The lateral displacement was calculated by comparing the impact point the muon would have if we extrapolated its momentum at production by a straight line trajectory with the actual impact point at ground.  The time delay was calculated by subtracting the known geometric and kinematic delays from the total delay given by CORSIKA. This {\it remaining} time delay includes the effects of the multiple scattering and also the differences between the real geometric and kinematic delays and the approximations made in the previous sections. The scattering angles were calculated by comparing the momentum at production with the momentum at ground, and will be used in \cite{a:MPDback}.

Fig. \ref{f:RMSy} displays the RMS of the lateral displacement as a function of the traveled depth for different  zenith angles with a final energy 1 and 10 GeV. The solid black line represents the analytic approximation given in \cite{a:PDG}. The effect of the multiple scattering on the position of the muons is accounted for in the model Monte Carlo as follows: the position of the muon is displaced from its geometric extrapolation following a 2-Gaussian distribution with RMS given by the look-up tables. The arrival time delay is also corrected by adding the extra term that accounts for the multiple scattering and the precision effects, sampled from a log-Gaussian whose parameters were also recorded in the look-up tables.

After the geomagnetic and multiple scattering correction, the total distance $l$ from the production point is reevaluated and the final energy is recalculated.  The effects of the multiple scattering on all the analyzed profiles, where the distances to the core are larger than 100 m, were found to be negligible. This is so because in most practical cases the width of the profile is wider than the width introduced by the multiple scattering smearing. Notice that when analyzing muon by muon these effects might become important in some cases.

\section{Ground distributions}

The model Monte Carlo transports ${\cal N}_0/w_i$ muons from the production point to the ground level. The weight at production $w_i$ can be selected as a trade off between speed and statistics (the present plots were done with $w_i=10$) and it is transformed at ground into a weight $w$, which is different for every muon, depending on its particular trajectory and energy. Weighted muons at ground were recorded and their distributions analyzed in comparison to the standard CORSIKA output.

\subsection{The energy distribution}
\label{s:ground_spectrum}
The energy at ground $E_f$ was histogramized and analyzed as a function of the impact point on ground $(r,\zeta)$. Typically, the muon energy is not directly measured by cosmic ray detectors since it would require extensive areas with particle detectors like those used in accelerator experiments and that is very expensive so far. Nevertheless, the spectrum of muons has an impact on other quantities that are measured by current air shower detector arrays, like the muon lateral distribution at ground, the arrival angle \cite{a:MPDback}, and the arrival time delay.

\begin{figure}[!h]
  \begin{center}$
\begin{array}{cc}
   \includegraphics[width=7cm]{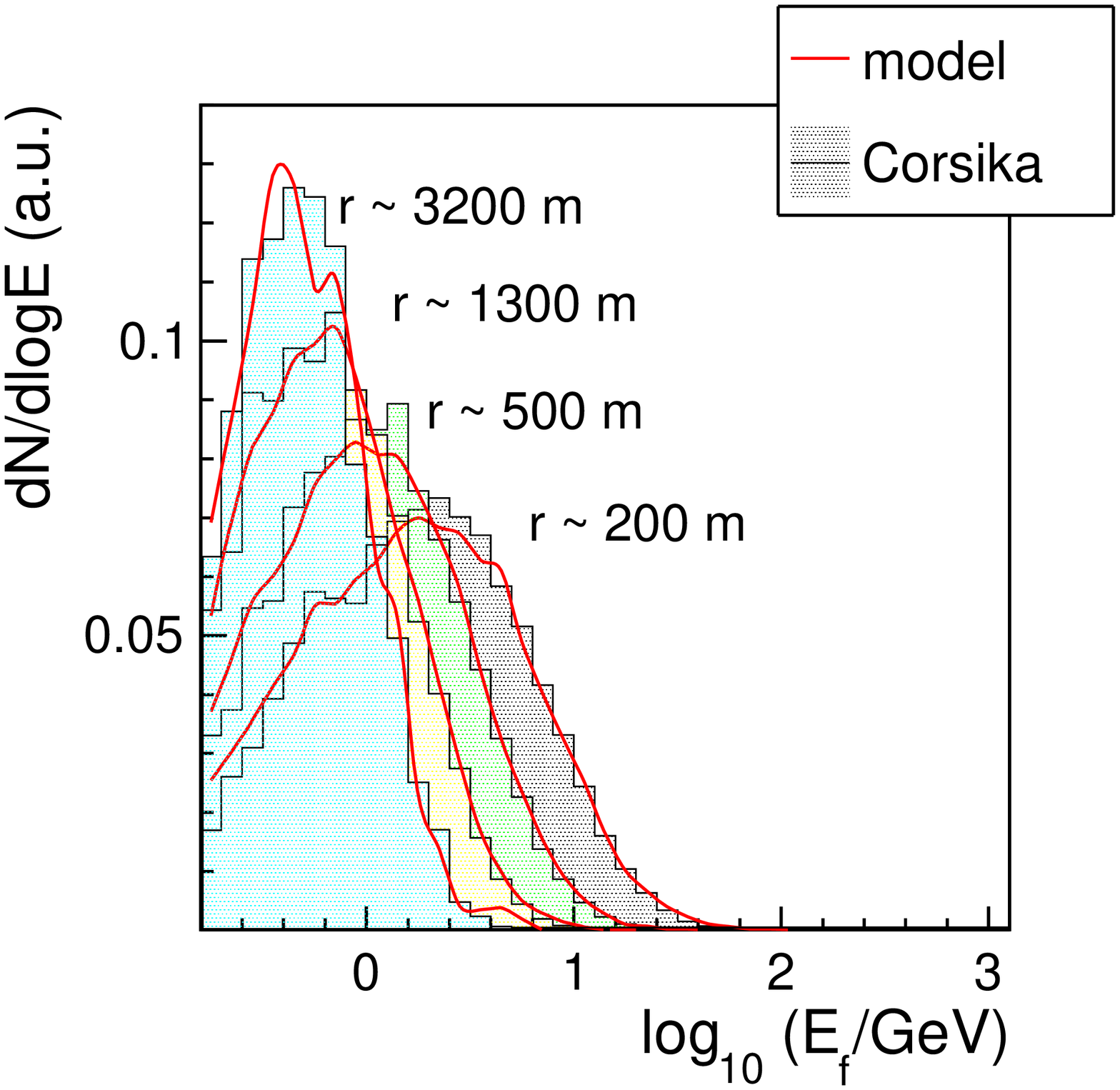}&
   \includegraphics[width=7cm]{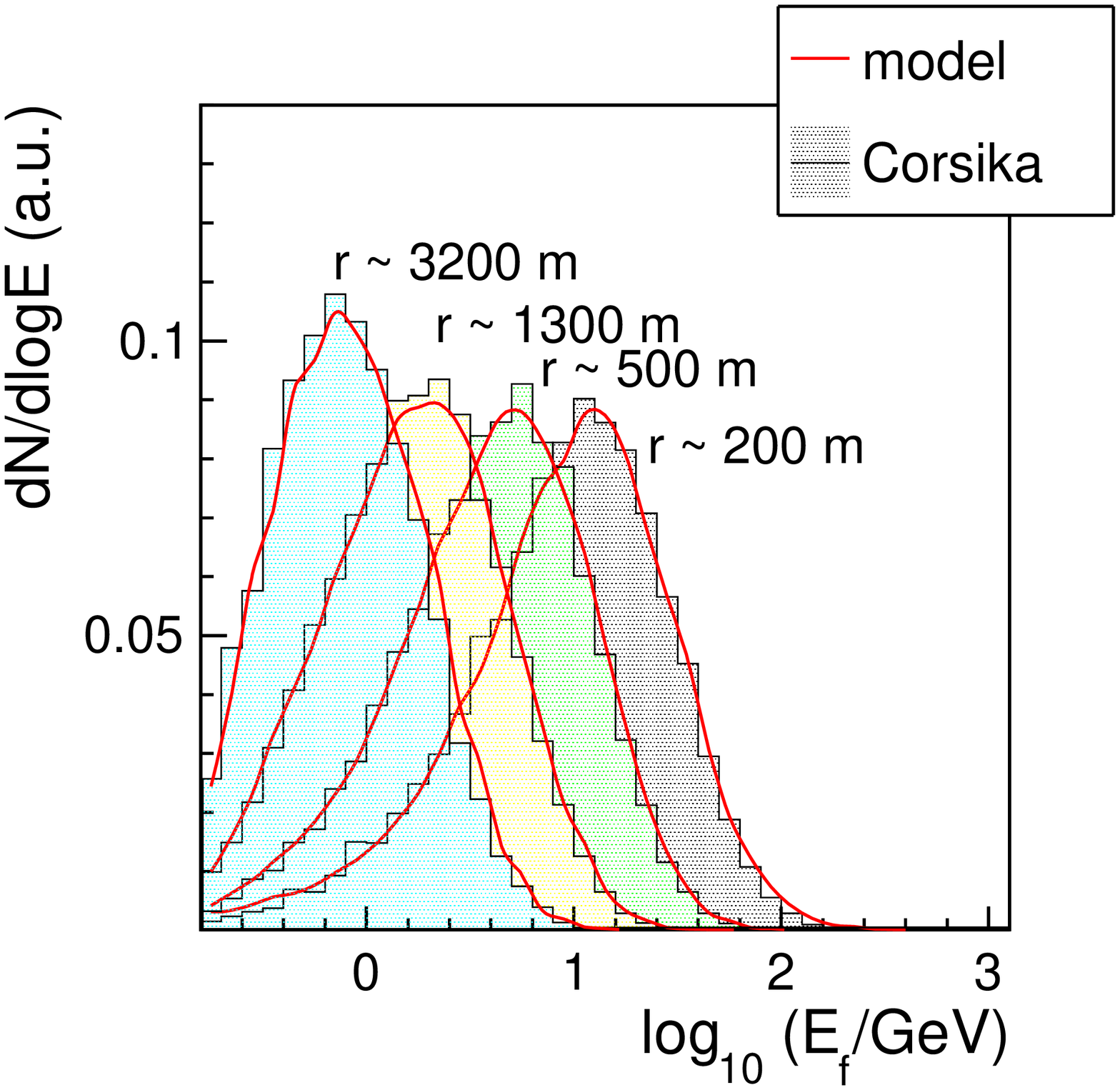}
\end{array}$
    \caption[]{Normalized energy spectrum of muons arriving at ground for a 0 deg shower (left panel) and for a 60 deg shower (right panel), at different distances from the core as given by corsika compared to the prediction of the model. }
    \label{f:comparisonlogE0deg}
  \end{center}
\end{figure}

Fig. \ref{f:comparisonlogE0deg} displays a comparison between CORSIKA and the model for the normalized energy spectra of a 0 deg and a 60 deg shower, at different distances from the shower core. The energy of muons decreases as $\sim 1/r$ and increases with the zenith angle \cite{a:TimeModel,a:PhDCazon}, being the details given by the $p_t$, $z$ and $E_i$ distributions. Low energy muons dominate at large distances from the core. 

\subsection{Apparent production depth distribution} The production depth of the detected muons, the {\it apparent} MPD-distribution, follow distributions whose shape changes with the observation coordinates. Muons can be understood as light rays coming out of the shower axis with an radiance which is not isotropic, following an angular distribution given by the $p_t$ and $E_i$ distributions as
\beq
\frac{d^4N}{dE_i d\Omega dX} \simeq \frac{1}{2 \pi} h(X) f_X(E_i,\sin \alpha E_i) E_i \frac{\cos \alpha}{\sin \alpha} 
\eeq
provided that the geomagnetic and multiple scattering effects are negligible.
Since the angle $\alpha$ subtended from the observation point $(r,\zeta)$ changes with the production point at the shower axis $z$, the number of muons observed to come from a particular $X$ will be different. Also, the production energy $E_i$ and the traveled distance $l$ will be different and will induce different decay probabilities, modifying the shape of the {\it total/true} muon production depth distribution through $p(E_i,l)$ into the {\it apparent} particular distribution of detected muons at that particular observation point.

\beq
{ 1 \over r}{d^3 N \over dX dr d\zeta}\simeq \int \frac{d^4N}{dE_i d\Omega dX} \frac{z}{l^3} p(E_i,l) dE_i
\eeq
We denote ${ 1 \over r}{d^3 N \over dX dr d\zeta}$ as $dN/dX|_{(r,\zeta)}$ for short.

\begin{figure}[!h]
  \begin{center}
    \includegraphics[width=10cm]{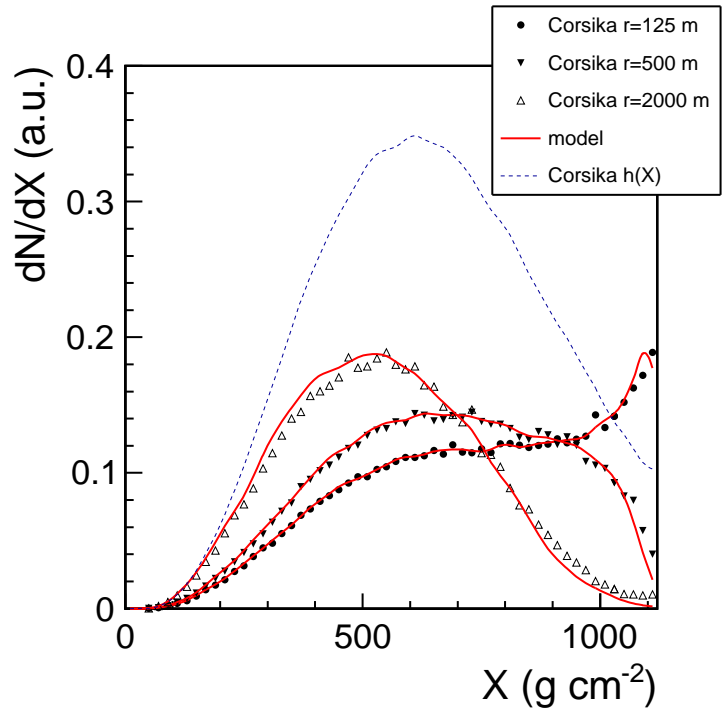}
    \caption[]{Comparison of several {\it apparent} MPD-distributions, $dN/dX|_{(r,\zeta)}$, for a 40 deg shower at different distances from the core. The {\it total/true} MPD-distribution ($h(X)$) is also plotted for comparison. Normalizations are arbitrary.}
    \label{f:dNdX.40}
  \end{center}
\end{figure}

\begin{figure}[!h]
  \begin{center}
    \includegraphics[height=7cm]{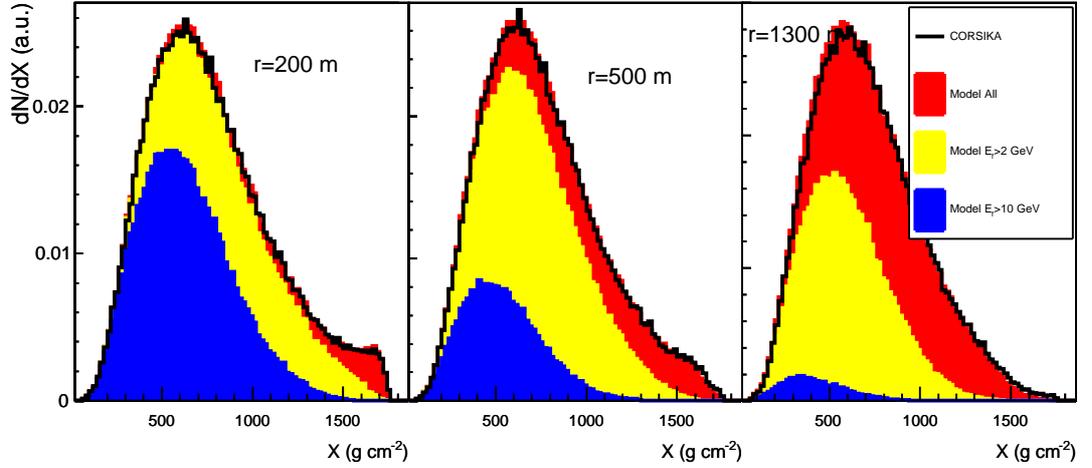}
    \caption[]{Comparison between model and CORSIKA of the normalized {\it apparent} MPD-distributions, $dN/dX|_{(r,\zeta)}$, for a 60 degree shower at 3 different distances from the core. The color histograms show the contribution of different energies.}
    \label{f:X.60}
  \end{center}
\end{figure}

Fig. \ref{f:dNdX.40} displays a comparison between CORSIKA and the present model of the {\it apparent} muon production depth distributions for a 40 deg shower at different distances from the core, where the distortions introduced in the $dN/dX|_{(r,\zeta)}$ distributions when compared to $h(X)$ can be observed.
Fig. \ref{f:X.60} displays a comparison at 60 degrees at different distances from the core. The distortions due to the observation point are small at high zenith angle. Different colors show the contributions from different energy of muons at ground. It can be seen that high energy muons tend to come from higher up in the atmosphere. At close distances to the core their contribution is enhanced.
 
The $dN/dX|_{(r,\zeta)}$ distribution is never directly observed, but reconstructed from the arrival time or the arrival angle at ground. The correct inference of the {\it total/true} MPD-distribution, $h(X)$, requires the knowledge of the exact dependence of $dN/dX|_{(r,\zeta)}$ with the observation point coordinates and detection energy threshold. $dN/dX|_{(r,\zeta)}$ explores different kinematic regions at production when reconstructed at different distances from the core. For instance, the algorithm proposed in \cite{a:MPD} and \cite{GarciaGamez:2011pe} requires the conversion of each $dN/dX|_{(r,\zeta)}$ observed in each station to an universal distribution in order to sum up the contributions of all detectors in a single shower.

\subsection{Time distributions}

\begin{figure}[!h]
  \begin{center}
    \includegraphics[height=7cm]{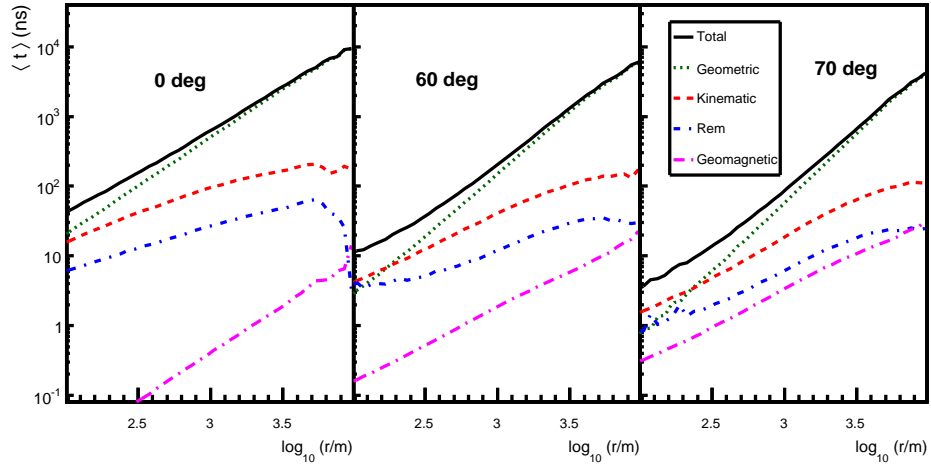}
    \caption[]{Average time delays for the different contributions to the total delay for 3 showers at $0$, $60$ and $70$ degrees, from left to right.}
    \label{delayscontribution}
  \end{center}
\end{figure}

\begin{figure}[!h]
  \begin{center}
    \includegraphics[height=7cm]{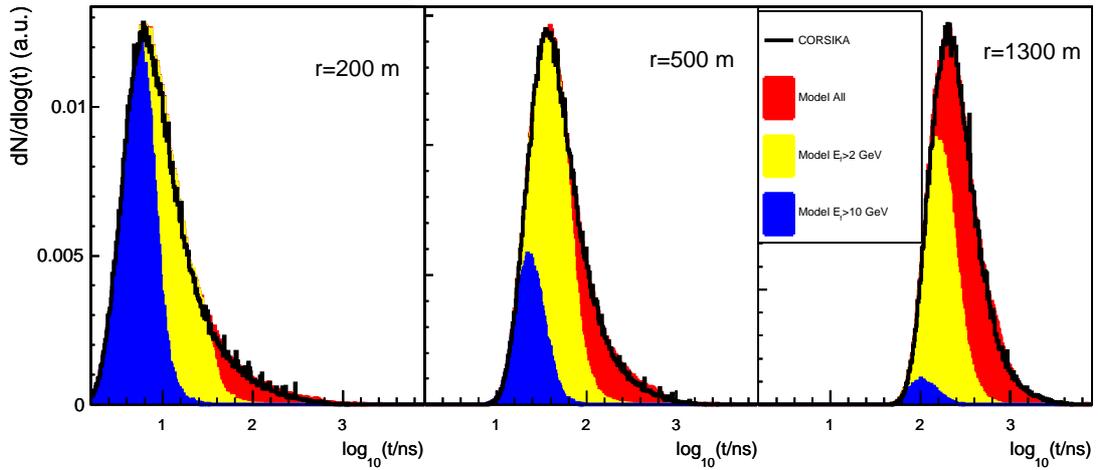}
    \caption[]{Comparison between the model and CORSIKA of the normalized time distributions for a 60 degree shower at 3 different distances from the core. The color histograms show the contribution of different energies.}
    \label{f:logt.60}
  \end{center}
\end{figure}

The total time delay is the sum of the different contributions calculated in the section 3. 
\beq
t=t_g+t_\epsilon+t_B+t_{Rem}
\eeq
where $t_g$ is the geometric delay, $t_\epsilon$ is the kinematic delay, $t_B$ is the contribution produced by the geomagnetic field, and finally $t_{Rem}$ accounts for the delay due to multiple scattering and the inacuracies due to the approximations used.
Fig. \ref{delayscontribution} displays the different delays for 3 different zenith angles, namely $0$, $60$ and $70$ degrees, where the different contributions to the total delay are displayed. The contribution to the multiple scatering is included in $t_{Rem}$.
At large distances from the core, the geometric delay is the most important. At distances typically from a few hundred meters to 1 km, the kinematic delay has a large impact. As we increase the zenith angle, the geometric delay looses importance relative to total delay. At $500$ m from the core, the geometric delay represents $\simeq$ 70\%,60\% and 50\% for 0, 60 and 70 degree showers, respectively. 
Fig. \ref{f:logt.60} displays the overall time distributions at 3 distances of the shower core for a 60 deg shower. Filled histograms show the contributions of different muon energies at ground. High energy muons arrive earlier at ground. This is so because they are produced higher up in the atmosphere, and therefore have less geometric delay, but also because they have less kinematic delay. 

The muon arrival time distributions can be used to extract important information. Far from the core, the time distributions are to a very good extent a one to one map of the {\it apparent} MPD-distributions. It can be determined by converting each muon time into a production distance, being the kinematic time a second order correction. Since the energy of each muon is typically not known, it is approximated by the mean value, taken from the energy spectrum at each observation point as discussed in section \ref{s:ground_spectrum} and as it was explained in \cite{a:MPD,a:TimeModel}. The energy would also determine the parameters of the multiple scattering delay distribution, although its concrete value follows a random distribution. The geomagnetic delay might take only two possible values depending on the charge of the muon. In general this technique will require a stringent $r$ cut for those regions where the geometric delay is a large fraction of the total delay, in order to avoid distortions of the reconstructed $dN/dX|_{(r,\zeta)}$. An alternative method consists in fitting the time distributions at once leaving a set of shape parameters on $h(X)$ free. A detailed discussion of this procedure will be made elsewhere.

\begin{figure}[!h]
  \begin{center} 
\includegraphics[height=7cm]{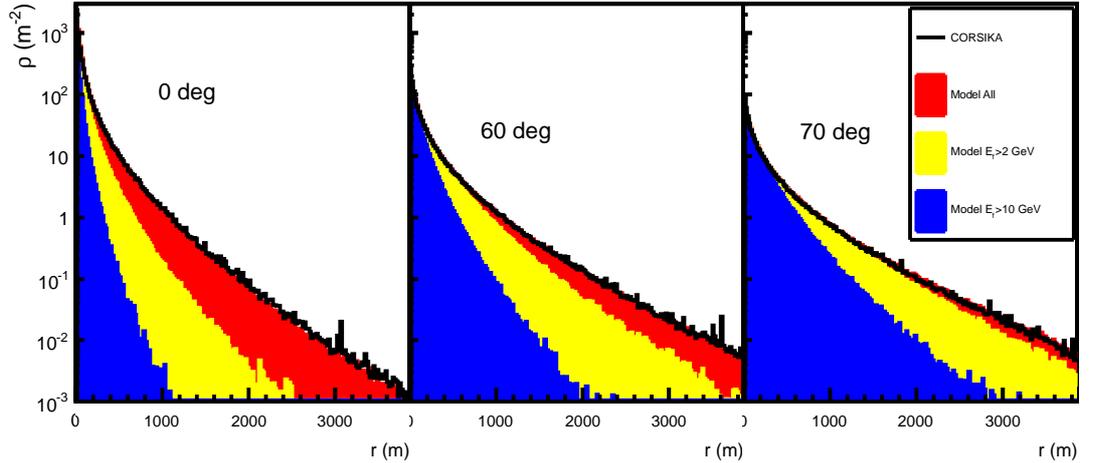}
    \caption[]{Comparison between the model and CORSIKA of the muon lateral distribution at ground for 3 showers at $0$, $60$ and $70$ degrees, from left to right. The color histograms show the contribution of different energies.}
    \label{f:r.60}
  \end{center}
\end{figure}

\begin{figure}[!h]
  \begin{center}
    \includegraphics[width=10cm]{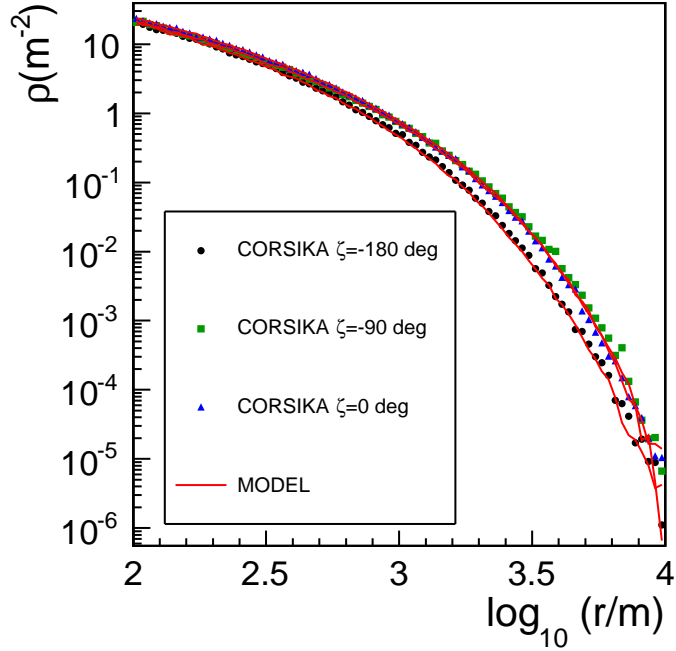}
\caption[]{Muon lateral distribution at ground for 3 different polar angles $\zeta$}
    \label{f:LDF036_70}
 \end{center}
\end{figure}

It is possible in principle to measure, or at least constrain, the shape of the muon spectrum. At distances close to the core the geometric delay is not dominant and the arrival time is mostly determined by the energy of each muon. A complete procedure to experimentally access this distribution is out of the scope of this paper, and will be analyzed elsewhere.

\subsection{Muon lateral distribution at ground}
The number of muons per surface area unit is $\rho(r,\zeta)=\frac{d^2N}{r dr d\zeta}$. Fig. \ref{f:r.60} displays examples of muon lateral distributions at ground for several zenith angles, where the contributions of different muon energies at ground were displayed in different colors. Low energy muons have a major impact on the fine details of the muon lateral distribution at ground.

 In vertical showers the number of muons per surface area does not depend much on $\zeta$. As we increase the zenith angle, asymmetries appear because of the different propagation effects, mainly decay and geometry.
The effects of the magnetic field become important above $60$ degrees, and they completely dominate the distributions at very inclined showers, typically between $80$ and $90$ degrees \cite{Ave:2000xs}. Fig. \ref{f:LDF036_70} displays the muon density as a function of $r$ for 3 different polar angles $\zeta$ on a 70 deg shower.

\vspace{1cm}

\begin{figure}[!h]
  \begin{center}$
  \begin{array}{cc}
    \includegraphics[width=7cm]{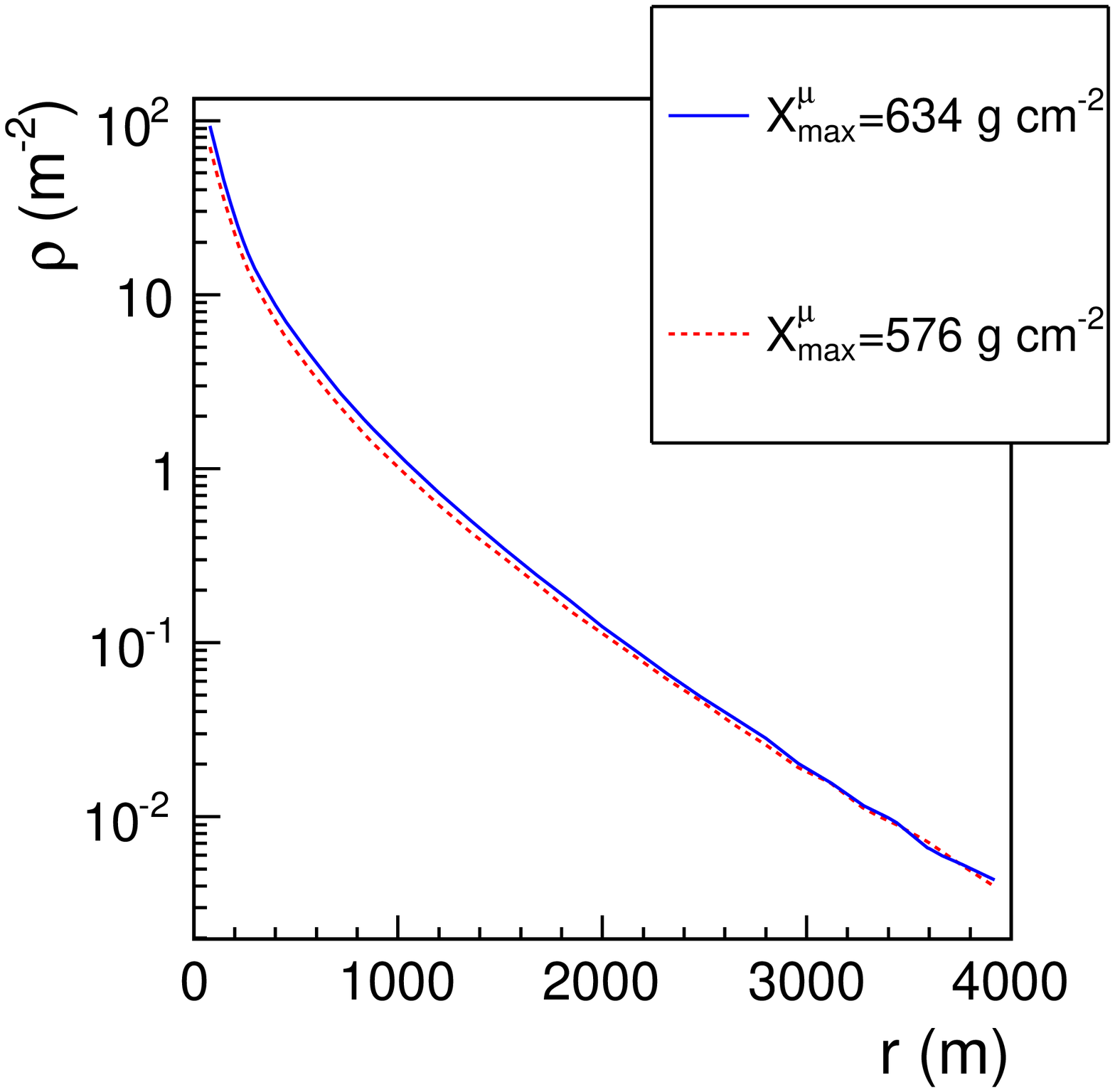}&
    \includegraphics[width=7cm]{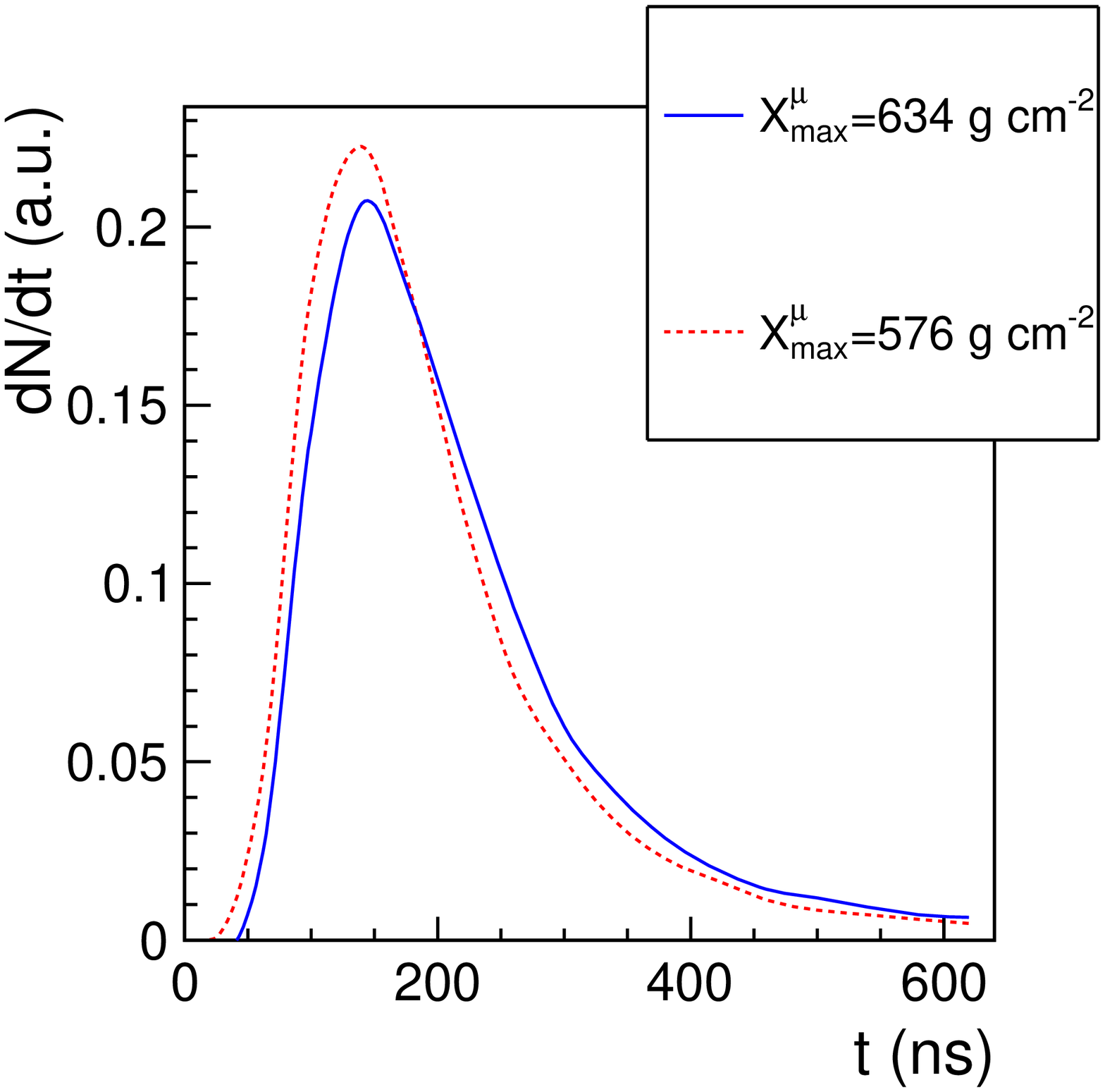} 
	\end{array}$
\caption[]{Muon lateral distribution (right panel) and normalized arrival time distribution at 2000 m and $\zeta=-180$ degrees (left panel)  for two shower maxima $\Xmax =634$ $\gcm$, and $\Xmax =576$ $\gcm$}
    \label{f:60.X634X576}
 \end{center}
\end{figure}

The shape of the ground distributions is fully determined by the distributions at production, $h(X)$ and $f_X(E_i,p_t)$.  A change in the overall muon content of the shower, ${\cal N}_0$, produces a change in the muon density at ground, and therefore in the normalization of all distributions. The other main source of fluctuations comes from the depth of the first interaction, which directly affects $h(X)$ by changing its maximum, $\Xmax$. The position of $\Xmax$  directly influences all distribution at ground since it changes the total distance traveled by muons to ground. Fig. \ref{f:60.X634X576} left panel displays two different muon lateral distributions at ground for two different positions of $\Xmax$ for a shower of $60$ degrees, where a change on the shape of the muon lateral distribution can be observed.  Fig. \ref{f:60.X634X576} right panel displays the normalized time distributions for $(r,\zeta)$=(1000 m,-180 deg), for the same showers.

\section{Average energy and transverse momentum distributions}

One of the main applications of the present model, is to be used in a global fit to extract information on the total number of muons in the shower ${\cal N}_0$, and the {\it total/true} production depth distribution, $h(X)$, and its maximum, $\Xmax$. In order to do so, we must assume a $f_X(E_i,p_t)$ distribution.

The energy and transverse momentum distributions display more universal features when they are expresed in terms of $X'=X-\Xmax$, once the effects of the fluctuations induced by the first interaction point are removed. Fig. \ref{f:hX_and_spectrum}, right panel, displays an example of the energy spectrum for 50 showers at $X'=-300$ $\gcm$ and $X'=300$ $\gcm$. For comparison, Fig. \ref{f:E_angles_Depths}, left panel, displays the averaged distributions over the $X'$ variable.  The average energy and transverse momentum distributions do not change when changing the energy of the primaries (Figs.\ref{f:E_Models} and \ref{f:pt_Models}, right panel), whereas they show mild differences between proton and iron primaries (Figs. \ref{f:E_primaries_angles} and \ref{f:pt_degs_primaries}, right panel), hadronic interaction models (Figs. \ref{f:E_Models} and \ref{f:pt_Models}, right panel).

If we substitute $f_X(E_i,cp_t)$ of a given shower by an average over showers of the same hadronic interaction model, primary, and zenith angle, $<f_{X'}(E_i,cp_t)>$, and leaving only $h(X)$ from the original shower, the ground density displays differences of about $\sim$ 2\% at 1000 m compared to the prediction it had occured if we used $f_X(E_i,cp_t)$, whereas the rest of the ground distributions remained unchanged.  It is thus possible to use an universal energy and traverse momentum distribution that depends only on $X'$, where the position of $\Xmax$ is naturally accounted for through $X=X'+\Xmax$.

 The systematics of any concrete aplication, including a global fit are to be studied and accounted for in each particular method and/or experimental setup. The effects of the choice of hadronic interaction model on $<f_{X'}(E_i,cp_t)>$ might introduce some systematics that should be also accounted for.
 
 One could also think of a method to experimentally constraint the energy and transverse momentum spectrum based on simultaneous observations of the ground distributions in different conditions.

\section{Conclusions}

In this paper we have built a model that explains the time delay, lateral distribution and depth profiles of muons at ground in terms of the distributions of energy, transverse momentum and production depth of all muons at the shower axis.  The propagation of muons takes into account, continuous energy losses, decay, magnetic field effects and multiple scattering. The effects of multiple scattering in the time distributions, the lateral distribution, {\it apparent} depth profile, and the energy spectrum can be neglected at distances to the core larger than 100 m. The angular distribution of muons at ground and the importance of the different effects will be analyzed elsewhere.

 This model can be used to experimentally reconstruct the distributions at production, in particular the {\it total/true} muon production depth $h(X)$ and the total number of muons ${\cal N}_0$. The energy and transverse momentum distributions at production play a fundamental role on shaping the time distributions and muon lateral distribution at ground. The $f_X(E_i,p_t)$ distribution shows a \emph{quasi}-universal behaviour when expressed in terms of $X'=X-\Xmax$ and it can substituted by an average of showers performed over the $X'$ variable in many applications.

 The model was implemented in a fast Monte Carlo, on two different steps: straight line propagation, and corrections, allowing the production of all distributions at ground starting from the distributions at production.

\section{Acknowledgments}
We would like to thank S. Andringa, C. Espirito-Santo and R. Engel for carefull reading of this manuscript and their constructive comments. We also thank P. Gon\c{c}alves, R. Ulrich and D. Garcia-Gamez for numerous discussions about different aspects related to the production and propagation of muons. This work is partially funded by Funda\c{c}\~{a}o para a Ci\^{e}ncia e Tecnologia (CERN/FP11633/2010 and SFRH/BPD/73270/2010), and fundings of MCTES through POPH-QREN Tipologia 4.2, Portugal, and European Social Fund.

\bibliographystyle{elsarticle-num}
\bibliography{<your-bib-database>}

\begin{thebibliography}{00}

\bibitem{Abraham:2010yv}
  J.~Abraham {\it et al.} [ Pierre Auger Observatory Collaboration ],
  Phys.\ Rev.\ Lett.\  {\bf 104 } (2010)  091101.
  [arXiv:1002.0699 [astro-ph.HE]].
\bibitem{Conceicao:2011vn}
  R.~Conceicao, J.~Dias de Deus and M.~Pimenta,
  Nuc.\ Phys.\ A (In Press)
 doi: 10.1016/j.nuclphysa.2012.02.019
 arXiv:1107.0912 [hep-ph].


\bibitem{d'Enterria:2011kw}
 D.~d'Enterria, R.~Engel, T.~Pierog, S.~Ostapchenko, K.~Werner,
 Astropart.\ Phys.\  {\bf 35 } (2011)  98-113.
 [arXiv:1101.5596 [astro-ph.HE]].

\bibitem{Tricomi:2010zz}
 A.~Tricomi, O.~Adriani, L.~Bonechi, M.~Bongi, G.~Castellini, R.~D'Alessandro, K.~Fukatsu, M.~Haguenauer {\it et al.},
 PoS {\bf ICHEP2010 } (2010)  026.

\bibitem{Ostapchenko:2005nj}
 S.~Ostapchenko,
 Phys.\ Rev.\  {\bf D74 } (2006)  014026.
 [arXiv:hep-ph/0505259 [hep-ph]].

\bibitem{Ostapchenko:2006vr}
 S.~Ostapchenko,
 Phys.\ Lett.\  {\bf B636 } (2006)  40-45.
 [hep-ph/0602139].



\bibitem{Allen:2011pe}
  J.~Allen for  The Pierre Auger Collaboration, 32nd International Cosmic Ray Conference, Beijing, China, August 2011
    [arXiv:1107.4804].

\bibitem{Grieder} Peter ~K. ~F. Grieder,  
Springer,(2010),ISBN 978-3-540-76940-8. 

\bibitem{Apel:2011zz}
  W.~D.~Apel, J.~C.~Arteaga, K.~Bekk, M.~Bertaina, J.~Blumer, H.~Bozdog, I.~M.~Brancus, P.~Buchholz {\it et al.},
  Astropart.\ Phys.\  {\bf 34 } (2011)  476-485.

\bibitem{Apel:2011mi}
  W.~D.~Apel {\it et al.}  [The KASCADE-Grande Collaboration],
  Phys.\ Rev.\ Lett.\  {\bf 107} (2011) 171104
  [arXiv:1107.5885 [astro-ph.HE]].

\bibitem{GarciaGamez:2011pe}
  D.~Garcia-Gamez for  The Pierre Auger Collaboration, 32nd International Cosmic Ray Conference, Beijing, China, August 2011
  [arXiv:1107.4804].

\bibitem{Cazon:2004zx}
  L.~Cazon, R.~A.~Vazquez, E.~Zas,
  Astropart.\ Phys.\  {\bf 23 } (2005)  393-409.
  [astro-ph/0412338].

\bibitem{Alfaro:2010zz}
  R.~Alfaro {\it et al.} [ Pierre Auger Collaboration ],
  Nucl.\ Instrum.\ Meth.\  {\bf A617 } (2010)  511-514.

 \bibitem{a:TimeModel}
  L.~Cazon, R.~A.~Vazquez, A.~A.~Watson, E.~Zas,
  Astropart.\ Phys.\  {\bf 21 } (2004)  71-86.
  [astro-ph/0311223].

\bibitem{a:MPD}
  L.~Cazon, R.~A.~Vazquez, E.~Zas,
  Astropart.\ Phys.\  {\bf 23 } (2005)  393-409.
  [astro-ph/0412338].

\bibitem{a:PhDCazon} L..~Cazon. PhD Thesis, Universidade de Santiago de Compostela, ISBN 84-9750-467-4


\bibitem{Bergmann:2006yz}
  T.~Bergmann {\it et al.},
  Astropart.\ Phys.\  {\bf 26} (2007) 420
  [arXiv:astro-ph/0606564].

\bibitem{CORSIKA}
Heck, D. et al
FZKA Report, ({\bf 6019}) (1998)

\bibitem{Ahn:2009wx}
  E.~J.~Ahn, R.~Engel, T.~K.~Gaisser, P.~Lipari and T.~Stanev,
  Phys.\ Rev.\  D {\bf 80} (2009) 094003
  [arXiv:0906.4113 [hep-ph]].

\bibitem{Werner:2005jf}
 K.~Werner, F.~-M.~Liu, T.~Pierog,
 Phys.\ Rev.\  {\bf C74 } (2006)  044902. [hep-ph/0506232].

\bibitem{Pierog:2006qv}
  T.~Pierog and K.~Werner,
  Phys.\ Rev.\ Lett.\  {\bf 101} (2008) 171101  [arXiv:astro-ph/0611311].



\bibitem{Schmidt:2007vq}
  F.~Schmidt, M.~Ave, L.~Cazon, A.~S.~Chou,
  Astropart.\ Phys.\  {\bf 29 } (2008)  355-365. [arXiv:0712.3750 [astro-ph]].



\bibitem{a:hXRuben} 
  S.~Andringa, L.~Cazon, R.~Conceicao and M.~Pimenta,
  Astropart.\ Phys.\  {\bf 35} (2012) 821
  [arXiv:1111.1424 [hep-ph]].

\bibitem{Ave:2000xs}
M. Ave, R.A. Vazquez, E. Zas, Published in Astropart.Phys.14:91,2000.
\bibitem{a:PDG}
  K.~Nakamura {\it et al.} [ Particle Data Group Collaboration ],
  J.\ Phys.\ G {\bf G37}, 075021 (2010). 
\bibitem{a:MPDback} L. Cazon {\it et al}, in preparation

\end{thebibliography}

\end{document}